\documentclass[trackchanges,twocolumn]{aastex7}
\usepackage{graphicx}
\usepackage{txfonts}
\usepackage{color}
\usepackage{comment}

\newcommand{\LIFE}{{\it LIFE}}

\begin{document}

\title{Probing thermal gradients of habitable-zone rocky planets as an anti-indicator of a global surface ocean\\using mid-infrared direct imaging}

\shorttitle{Probing thermal gradient with direct imaging}
\shortauthors{Fujii et al.}

\author[0000-0002-2786-0786]{Yuka Fujii}\affiliation{Division of Science, National Astronomical Observatory of Japan, 2-21-1 Osawa, Mitaka, Tokyo 181-8588, Japan}
\affiliation{Graduate Institute for Advanced Studies, SOKENDAI, 2-21-1 Osawa, Mitaka, Tokyo 181-8588, Japan}
\affiliation{Department of Earth and Planetary Science, University of Tokyo, 7-3-1 Hongo, Bunkyo-ku, Tokyo 113-0033, Japan}
\email{yuka.fujii.ebihara@gmail.com}

\author[orcid=0000-0001-6138-8633]{Daniel Angerhausen} 
\affiliation{ETH Zurich, Institute for Particle Physics \& Astrophysics, Wolfgang-Pauli-Str. 27, 8093 Zurich, Switzerland}
\affiliation{Blue Marble Space Institute of Science, Seattle, WA, USA}
\affiliation{SETI Institute, 189 N. Bernado Ave, Mountain View, CA 94043, USA}
\email{dangerhau@phys.ethz.ch}

\author[orcid=0009-0003-7304-7512]{Taro Matsuo}
\affiliation{Department of Earth and Space Sciences, Graduate School of Science, University of Osaka}
\email{matsuo@ess.sci.osaka-u.ac.jp}

\author[orcid=0000-0002-7188-1648]{Eric T. Wolf}
\affiliation{Laboratory for Atmospheric and Space Physics, University of Colorado Boulder}
\email{eric.wolf@colorado.edu}


 
\begin{abstract}
Future direct-imaging missions such as the Large Interferometer for Exoplanets (\LIFE{}) aim to observe thermal emission from potentially habitable planets to characterize their surface environments and search for signs of life. 
Previous studies of directly imaged Earth-like planets have mainly examined the signatures of atmospheric composition, 
often using one-dimensional models, while the effect of horizontal temperature gradients has received limited attention. 
Because a pronounced horizontal temperature gradient may signal the absence of a global ocean, we investigate its detectability through thermal-emission direct imaging. 
Adopting Teegarden's Star b (zero-albedo equilibrium temperature $\sim 280$~K) as a benchmark, we compute three-dimensional atmospheric structures with and without a global ocean using the ROCKE-3D general circulation model and simulate geometry-dependent thermal emission spectra. 
We show that the temperature gradients that disfavor a global-ocean scenario manifest in both orbital phase variation and spectral shape of the snapshot spectra. The phase variation is more readily detectable: one-day integrations with \LIFE{} at two orbital phases would reveal flux variations in no-ocean cases with 1-10~bar atmospheres, depending on background atmospheric composition. 
Shapshot spectra provide complementary diagnostics, including the running brightness temperature of the continuum and detailed absorption band shapes, 
but require longer integration times. 
These three-dimensional effects, if neglected, can bias interpretations based on one-dimensional models, and could also lead to misidentification of the H$_2$O vapor band at wavelengths beyond $\sim$10~$\mu$m. 
We also assess the detectability of these 3D effects on other nearby exoplanets. 
Our results highlight the need to incorporate three-dimensional atmospheric structures when characterizing rocky exoplanets, both to constrain surface conditions and to avoid misinterpretation of spectral data. 
\end{abstract}

\keywords{\uat{Exoplanet atmospheres}{487} --- \uat{Exoplanet surface characteristics}{496} --- \uat{Extrasolar rocky planets}{511} --- \uat{Habitable planets}{695} --- \uat{Exoplanet atmospheric dynamics}{2307}}
%
\section{Introduction}

Characterizing Earth-sized exoplanets in the so-called habitable zones (hereinafter potentially habitable planets) has been one of the major goals of exoplanet observations. 
While the targets of the James Webb Space Telescope (JWST) are approaching this parameter space, major advances in large-scale surveys will be enabled by the capability of direct imaging observations with efficient starlight suppression technologies. 
Direct imaging of potentially habitable planets are planned both in shortwave (the UV/optical/near-infrared), where planets are illuminated predominantly by scattering the stellar light, and in longwave, where planets emit by their own thermal emission. 
NASA's space-based observatory, the Habitable World Observatory \citep[HWO;][]{2021pdaa.book.....N}, aims at characterizing potentially habitable planets in shortwave, while the Large Interferometer For Exoplanets \citep[\LIFE{};][]{LIFE1}, a space-based infrared nulling interferometer, would image planets in longwave. 
The next-generation 30-40-meter class ground-based telescopes are also
planned to be equipped with coronagraphic instruments to enable direct imaging in both shortwave and longwave. 
While these different facilities are expected to provide complementary information \citep{Fujii+2018}, observing planetary thermal emissions in longwave is particularly useful for constraining thermal structures as well as detecting fundamental rot-vibrational bands of atmospheric compositions.
Whether these diagnostics can ultimately identify habitable (and inhabited) conditions is a fundamental question. 

To gain insights into the properties of thermal emission spectra of habitable planets, 
the thermal emission of the Earth has been studied in detail.  
Earth's thermal spectra show prominent bands of CO$_2$, O$_3$, H$_2$O, and CH$_4$ \citep{Tinetti+2006a,Tinetti+2006b,Robinson+2011}, regardless of season and viewing geometry, while the continuum flux and absorption band strengths vary by no more than 20\% in most cases, as quantified by the ratio of the variation amplitude to the mean \citep{GomezLeal+2012,Mettler+2020,Mettler+2023}.
The thermal emission spectra of various Earth-like planet scenarios have also been modeled, using either prescribed atmospheric profiles or self-consistently calculated profiles with radiative transfer and photochemistry under a range of boundary conditions
\citep{KalteneggerSasselov2010,Kaltenegger+2010,Rugheimer+2013,Rugheimer+2015,RugheimerKaltenegger2018,Arney+2016,Braam+2025}. 
Furthermore, with the emergence of the \LIFE{} mission concept, simulated observations of modeled thermal emission have been generated assuming a \LIFE{}-like capabilities, accompanied by assessments of data interpretation \citep[e.g.,][Rugheimer et al., submitted]{LIFE2,LIFE3,LIFE4,LIFE5,Angerhausen+2024,Taysum+2024,Konrad+2024,Boukrouche+2024}. 
Except for the Earth case, most of these spectral studies that specifically evaluate the capabilities of direct imaging observations have employed one-dimensional (1D) atmospheric models. 
While successful identifications of atmospheric species, including biosignatures candidates have been demonstrated, it remains elusive whether thermal emission spectra obtained with direct imaging can  constrain the presence of surface liquid water as well.

In the meanwhile, three-dimensional (3D) General Circulation Models (GCMs) have been used to  investigate the effect of atmospheric and surface conditions on the global atmospheric structure and their associated thermal emissions. 
Although the 3D atmospheric structures depend on myriad parameters, the presence of global surface liquid water has a fundamental role \citep[][]{Abe+2011,Kodama+2018}. 
Assuming synchronously rotating planets around low-mass stars, \citet{Yang+2013} calculated the 3D atmospheric structures with a GCM, and demonstrated that the top-of-atmosphere thermal emission of ocean-covered planets is horizontally more uniform than in dry rocky planets, due to the combined effects of the enhanced heat re-distribution and the optically thick cloud decks near the substellar points. 
These effects are most prominent near the inner edge of habitable zones, because of the increased abundance of water vapor in the atmosphere \citep{Yang+2019,Wolf+2019, KomacekAbbot2019}. 
Even for cooler planets, surface ocean contributes to efficient heat transport, compared to the case without ocean, through the phase changes of water as well as the ocean circulation \citep{Turbet+2016,DelGenio+2019}. 
The efficient heat transport due to the presence of surface water also applies to rapidly rotating planets. 
Based on the parameter studies of the atmospheric structures of ocean-covered rocky planets with a GCM, \citet{KaspiShowman2015} showed that meridional moist energy flux is comparable to the static energy flux, and becomes larger for warmer planets. 
These properties suggest that the temperature gradient can serve as an important constraints on the presence of global surface liquid water.

Clearly, there are also other factors that act to homogenize the top-of-atmosphere thermal emission. 
As the atmospheric thickness increases, horizontal heat transport through circulation becomes more efficient and the day-night temperature contrast is reduced \citep[e.g.,][]{Selsis+2011,KollAbbot2015,Turbet+2016,Turbet+2018}. 
The efficiency of heat transport also increases as the abundance of radiatively active species increases in the atmosphere \citep{WangYang2022}. 
Planets globally covered with clouds, such as Venus, are another possibility. 
Thus, rather than uniform thermal emission as definitive evidence of a global ocean, one may argue that the presence of large temperature contrast provides evidence for its {\it absence} (i.e., anti-indicator), together with the absence of global cloud cover. 
While not conclusive, this may represent a unique constraint that thermal emission observation can provide, complementary to the constraints on the presence of liquid water from scattered-light observations \citep[see][for a review]{Robinson2018}. 

A number of previous GCM studies have produced planetary thermal emission associated with different climate states. 
The primary observational technique considered has been thermal phase curves, that is, disk-averaged planetary thermal emission as a function of planetary orbital (or rotational) phase, which is observationally obtained 
as the time-varying component of the combined star-planet spectra.
Thermal phase curves have been successfully detected for close-in transiting hot Jupiters \citep[e.g.,][]{Knutson+2007} and hot rocky planets \citep[e.g.,][]{Kreidberg+2019}. 
However, detecting such signals for  potentially habitable planets may be challenging even with JWST. 
While this technique is, in principle, applicable to both transiting and non-transiting planets, 
non-transiting planets are subject to uncertainty in the stellar flux levels, 
orbital inclinations, and planetary radii.  
Transiting planets, on the other hand, are intrinsically rare and only a small number of nearby targets are close enough to Earth to yield sufficient signal-to-noise ratio. 
These limitations motivate revisiting the 3D effects in the context of direct imaging missions, as a next-generation technique to characterize thermal emission. 

Building on this perspective, this paper aims to evaluate the potential of future direct imaging missions for constraining thermal gradient of potentially habitable planets as a diagnostic of the global oceans. 
Our primary focus is the \LIFE{} mission, which includes potentially habitable planets around nearby M-type stars as part of its observational targets \citep{LIFE1}.
We adopt Teegarden's Star b as a benchmark planet, and calculate 3D atmospheric structures for dry, moist, and ocean-covered scenarios, highlighting the difference in temperature gradients. 
These structures are subsequently used as input for mock observations with \LIFE{}. 
We consider two approaches to constraining temperature gradients through direct imaging. 
The first is the dependence of the flux on orbital phase (i.e., phase variation), analogous to thermal phase curve observations without direct imaging, and likely the easiest to constrain unless the orbit is nearly face-on. 
The second is to investigate the imprint of 3D thermal structures on single-epoch (``snapshot'') spectra. 
This approach is more closely aligned with previous studies of direct-imaging spectra and, in principle, applies to any target regardless of orbital inclination.
We compare spectra derived from 3D structures with those from conventional 1D profiles, pointing out observational signatures of global temperature gradients. 

The rest of this paper is organized as follows:
Section \ref{s:method} describes our method for  simulating 3D atmospheric structures with the GCM ROCKE-3D, calculating the resultant disk-averaged thermal emission spectra, and estimating the signal-to-noise ratio. 
The assumptions for the planetary parameters are also described there. 
Section \ref{s:result} presents the atmospheric structures obtained with ROCKE-3D, together with their associated thermal emission. 
The characteristics of the phase variation and snapshot spectra as the clues to horizontal temperature gradients are discussed, along with the assessment of their detectability. 
Section \ref{s:observability} extends the discussion to other targets and estimates the yield of the \LIFE{} mission. 
Section \ref{s:discussion} discusses the appropriate targets of the investigation, expanding on the dependence on the incident flux and the host star type, and how the interpretation of the spectra may be biased by the 3D effects. 
Section \ref{s:summary} summarizes our findings. 

\section{Method}
\label{s:method}

\subsection{Benchmark target: Teegarden's Star b}
\label{ss:method_target}

As a benchmark target, we have chosen Teegarden's Star b \citep{Zechmeister+2019}, one of the nearest potentially habitable planets, for the following reasons.  
Its close orbit around a nearby low-mass M7-type star ensures a reasonable integration time for characterization. 
It also increases the likelihood of synchronous rotation, which simplifies the interpretation of planetary thermal emission spectra and phase variations.  
Moreover, it orbits near the inner edge of the habitable zone, making it a favorable case for testing the ocean-world hypothesis, as discussed above (see also Section \ref{sss:dep_Sx}). 
Conveniently, the small inner working angle (IWA) of \LIFE{} \citep[10-20~mas;][see also Section \ref{ss:LIFESim} below]{KammererQuanz2018,Quanz+2018,Kammerer+2022} allows direct detection of thermal emission from Teegarden's Star b. 

In the climate modeling described below, we adopt system parameters based on \citet{Dreizler+2024}.  
The planet's true mass is assumed to be $M_{\rm p}=1.34M_{\oplus }$, inferred from the observed minimum mass $M_{\rm p}\sin i = 1.16M_{\oplus}$ assuming an inclination of $i = 60^{\circ}$.  
The planetary radius is estimated to be 1.1$R_{\oplus}$, based on the mass–radius relationship for Earth-like rocky planets from \citet{Zeng+2016}. 
The orbital period is 4.91 days, and under the assumption of tidal lock, the rotation period is identical to the orbital period. 
The obliquity is set to zero. 
The spectrally integrated incident flux is calculated to be $1.08\,S_{\oplus}$, where $S_{\oplus}$ is the solar constant, based on the stellar luminosity ($L_{\star} = 10^{-3.14}\,L_{\odot}$) and the semi-major axis ($a = 0.0259$~au). 
Although the latest analysis suggests a possible non-zero eccentricity ($e = 0.03^{+0.04}_{-0.02}$), we assume zero eccentricity for simplicity. 
If $e=0.02$, the ratio of incident flux between perihelion and aphelion would be 1.13, which could slightly influence the climate. 
Another Earth-sized planet, Teegarden's Star c, has been detected in this system \citep{Zechmeister+2019}; However, it is not included in our simulation.

%
\begin{table*}
\centering 
\caption{Assumptions for GCM calculations}             
\label{tbl:models}      
\begin{tabular}{l l l l l}        
\hline\hline                 
Name & Surface & Atmospheric composition & Atmospheric pressure & Note \\    
\hline                        
\texttt{ocean\_Nc-6} & zero-flux ocean of 50~m depth & N$_2$-dominated with 1~ppm CO$_2$ & \{1~bar, 10~bar\} & \\      
\texttt{dry\_Nc-6} & dry sand & N$_2$-dominated with 1~ppm CO$_2$   & \{1~bar, 10~bar\} & \\
\texttt{dry\_Nc-4} & dry sand & N$_2$-dominated with 100~ppm CO$_2$   & \{1~bar, 10~bar\} & \\
\texttt{dry\_Nc-2} & dry sand & N$_2$-dominated with 1\% CO$_2$   & \{1~bar, 10~bar\} & \\
\texttt{dry\_C}    & dry sand & 100\% CO$_2$   & \{1~bar, 10~bar\} \\
\texttt{moist\_Nc-6} & sand & N$_2$-dominated with 1~ppm CO$_2$ $+$ 1\% H$_2$O vapor & 1~bar & hot-start$^{*}$ \\
\texttt{moist\_Nc-4} & sand & N$_2$-dominated with 100~ppm CO$_2$ $+$ 1\% H$_2$O vapor & 1~bar & hot-start$^{*}$ \\
\texttt{moist\_Nc-2} & sand & N$_2$-dominated with 1\% CO$_2$ $+$ 1\% H$_2$O vapor,    & 1~bar & hot-start$^{*}$ \\
\texttt{moist\_C}    & sand & 100\% CO$_2$ $+$ 1\% H$_2$O vapor & 1~bar & hot-start$^{*}$\\
\hline                                   
\end{tabular}
\begin{flushleft}
$^{*}$The temperature is initialized with 400~K everywhere and all the water are initially in the vapor phase. 
\end{flushleft}
\end{table*}
%

\subsection{Scenarios for atmospheres and surfaces}
\label{ss:scenarios}

We study the 3D atmospheric structures of the benchmark planet Teegarden's Star b. 
While there have already been many studies on the climate of habitable (ocean-covered) planets around low-mass stars as well as the associated thermal phase curves, 
this study focuses on the difference between globally ocean-covered and no-ocean scenarios. 
Thus, we consider three classes of scenarios: one with a global ocean, and two without an ocean, differing in the presence or absence of atmospheric water vapor. 
The setups for different scenarios of these three scenarios are described below and summarized in Table \ref{tbl:models}. 

\vspace{0.5\baselineskip}

{\bf Globally ocean-covered scenarios}  (\texttt{ocean\_Nc-6}). 
This model assumes that the surface is globally covered with an ocean without a land. 
It is a thermodynamic (``q-flux'') ocean with 50~m depth \citep[e.g.,][]{Godolt+2015,Fujii+2017}, without lateral ocean heat transport (hence ``zero-flux'' in Table \ref{tbl:models}). 
Ignoring the ocean heat transport has only a minor effect on the heat transfer of planets near the inner edge of habitable zones, due to the substantial heat transport by atmospheric water vapor \citep{Yang+2019}. 
Note that the possible presence of even sparse continents could disrupt day-night ocean heat transport patterns \citep{Yang+2019,DelGenio+2019}, thereby further reducing the impact of ocean heat transport. 
The atmosphere is assumed to be N$_2$-dominated, with 1~ppm CO$_2$ volume mixing ratio. 
The atmospheric pressure is set to 1 bar or 10 bar. 

\vspace{0.5\baselineskip}

{\bf No-ocean dry scenarios} (\texttt{dry\_*}). 

The planet is covered with sand. 
Neither surface nor atmospheric moisture is included.
The atmosphere is assumed to be composed of N$_2$ and CO$_2$, and the CO$_2$ volume mixing ratio is varied among 1~ppm, 100~ppm, 1\%, and 100\%, based on the insight that the large mixing ratio of CO$_2$ increases the heat transport efficiency \citep{WangYang2022}. 
As the heat transport efficiency also depends on the atmospheric thickness, the atmospheric pressure is set to 1 bar or 10 bar.

\vspace{0.5\baselineskip}

{\bf No-ocean moist scenarios} (\texttt{moist\_*}). 

This class represents the intermediate case between the two scenarios described above. 
The planet lacks a surface ocean and is instead covered with sand, but its atmosphere contains moisture. 
The model is initialized 
with a water vapor mass mixing ratio (or specific humidity, $q$) of $10^{-2}$ and atmospheric temperature of 400~K everywhere (i.e., hot-start), and is evolved until it reaches a steady state. 
The atmospheric water abundance in the steady state is smaller than initially assumed due to the condensation onto the surface, to the varying degree depending on the background atmospheric composition. 
Similar to the \texttt{dry\_*} runs, the background atmosphere is composed of N$_2$ and CO$_2$ with varying CO$_2$ mixing ratios, while the atmospheric pressure is set to 1~bar.

\subsection{Calculating 3D atmospheric structures and thermal emission spectra with ROCKE-3D}
\label{ss:ROCKE3D}

We simulate the 3D atmospheric structures of the above scenarios for the benchmark target Teegarden's Star b employing the ROCKE-3D\footnote{https://simplex.giss.nasa.gov/gcm/ROCKE-3D/} (Resolving Orbital and Climate Keys of Earth and Extraterrestrial Environments with Dynamics) GCM \citep{Way+2017}. 
It was developed from the NASA Goddard Institute for Space Studies (GISS) ModelE2 (Schmidt et al. 2014), and has been applied to various solar and extrasolar planets \citep[e.g.,][]{Way+2016,Fujii+2017,DelGenio+2019,Colose+2021}. 
The ROCKE-3D version used in this work is \texttt{planet\_1.1}. 

The horizontal grid is set to $4^{\circ }\times 5^{\circ }$ latitude-longitude, 
and the atmosphere has 40 vertical layers up to 0.1 mbar, consistent with previous runs for tidally locked potentially habitable planets around M-type stars \citep{Fujii+2017}. 

The radiative transfer is solved using SOCRATES \citep{Edwards1996, EdwardsSlingo1996,Amundsen+2017}. 
The opacities of CO$_2$ and H$_2$O are based on HITRAN2012 \citep{Rothman+2013} and are treated using the correlated-k method \citep{Goody+1989,LacisOinas1991}, where k-terms are derived using exponential sum-fitting of transmissions \citep{WiscombeEvans1977}.
The number of bands used in radiative transfer is chosen to ensure adequate accuracy, depending on the atmospheric composition, as follows. 
We adopted 15 shortwave bands and 43 longwave bands except for CO$_2$ 100\% runs. 
For CO$_2$ 100\% runs, we adopted 17 shortwave bands and 46 longwave runs. 
The stellar spectral energy distribution is derived by interpolating the BT\_Settl model \citep{Allard+2012} to $T_{\rm eff} = 2904$~K, [Fe/H]$=-0.19$, and $\log g = 5.32880$. The stellar temperature assumed here is based on an earlier measurement of \citet{Zechmeister+2019}, reflecting the timing of the project’s initiation. 
Although these parameters differ slightly from the more recent values reported by \citet{Dreizler+2024}, the differences are unlikely to substantially affect our conclusions.

We note that the HITRAN2012 database employed by the radiation module of \texttt{planet\_1.1} is validated for the temperature range between 100 and 400 K.  
As we will see in Section \ref{s:result}, our hottest runs slightly violate these conditions; the maximum temperatures of \texttt{dry\_C\_10bar} and \texttt{moist\_C\_1bar} are 404~K and 419~K. 
Thus, caution is warranted when referring to the quantitative results of the hottest runs.

Ocean albedo in ROCKE-3D is calculated based on water and sea foam reflectance \citep{Way+2017}, where water albedo is calculated as a function of the solar zenith angle and wind speed and the sea foam reflectance is derived from \citet{Frouin+1996}. 
Specifically, the albedo is several percent in the substellar regions (low solar zenith angle) and increases to the day-night terminator up to 60\%. 
The albedo for sand is assumed to be 0.14 and 0.0 for shortwave and longwave, respectively. While the shortwave surface albedo slightly affects the global temperature, the spectral shape of the longwave  albedo directly affects the thermal emission spectra when the atmosphere is optically thin. This albedo assumption is a simplification of real rocky surfaces, which would have wavelength-dependent emissivity, so that the effects of temperature structure are isolated. 
The shortwave albedo lies near the lower end of the albedo range for rocky surfaces \citep{Hu+2012}, and is consistent with some basaltic surfaces of Solar System rocky bodies such as the Moon and Mercury \citep{Fujii+2014}. 
The relatively low land albedo also facilitates comparison with ocean-covered scenarios by minimizing albedo differences. 

We ran each simulation until the atmospheric structure reached steady state, where the 50-orbit average of the planetary net radiation is within $\pm 1~\mathrm{W\,m^{-2}}$. 
We also tracked 
the evolution of the globally averaged surface temperature and ice cover fraction for the ocean-scenario runs (\texttt{ocean\_Nc-6}) to check whether the system reached a steady state. 
After each model reached a steady state, 
we resumed the GCM simulation using an opacity table with a finer wavelength grid, so that the thermal emission at each of these narrower spectral bins is calculated within the framework of the GCM and stored in the standard output. 
The spectral resolution of this finer wavelength grid is  $\Delta \lambda = 0.1\cdot (\lambda /10~\mu{\rm m})^{1/2}$, which corresponds to $\mathcal{R}\sim 100$ at $\lambda =10~\mu {\rm m}$, comparable to the spectral resolution of \LIFE{} assumed here ($\mathcal{R}=50$). 
To obtain the disk-averaged radiance of the planet, 
we averaged the thermal emission from surface patches with the weight of the projected area, i.e., $\Delta \Omega \cdot \cos \theta_{\rm obs}$, where $\Delta \Omega $ is the area of the patch, and $\theta _{\rm obs}$ is the emission angle (i.e., the angle between the normal of the patch and the observer's line of sight). 
This treatment assumes that the thermal emission from the top of the atmosphere (TOA) of each patch is isotropic. 
Here, the emission angle ($\theta _{\rm obs}$) of each surface patch is calculated as a function of the orbital inclination, $i$, and the orbital longitude, $\phi $, which is measured from the point of maximum dayside visibility. 
In other words, in the case of an edge-on orbit, $\phi =0^{\circ}$ and $\phi=180^{\circ}$ correspond to the eclipse and the transit, respectively, and $\phi = \pm 90^{\circ}$ correspond to the observations from above the western and eastern terminators relative to the substellar point.
These angles are related to the phase angle, $\alpha $, through the relation $\cos \alpha = \cos \phi \sin i$.

\subsection{Estimating Observability with \LIFE{}\textsc{sim}}
\label{ss:LIFESim}

\begin{table}[bh!]
\caption{Simulation parameters used in \LIFE{}\textsc{sim} in this study, which is in accordance with the parameter sets used in \cite{LIFE1} or \cite{LIFE8}}
\centering
\begin{tabular}{l l}
\hline\hline
Parameter & Value \\
\hline
     Quantum efficiency & 0.7\\
     Throughput & 0.05\\
     Minimum Wavelength & 4 $\mu$m       \\
     Maximum Wavelength & 18.5 $\mu$m   \\
     Spectral Resolution & 50      \\
     Nulling Baseline & 10-100 m \\
     Apertures Diameter & 2 m \\
     Exozodi & 3x local zodi \\
\hline\hline
\end{tabular}\label{tab:lifesim}
\end{table}

\begin{figure*}[bt!]
\centering
\includegraphics[width=\hsize]{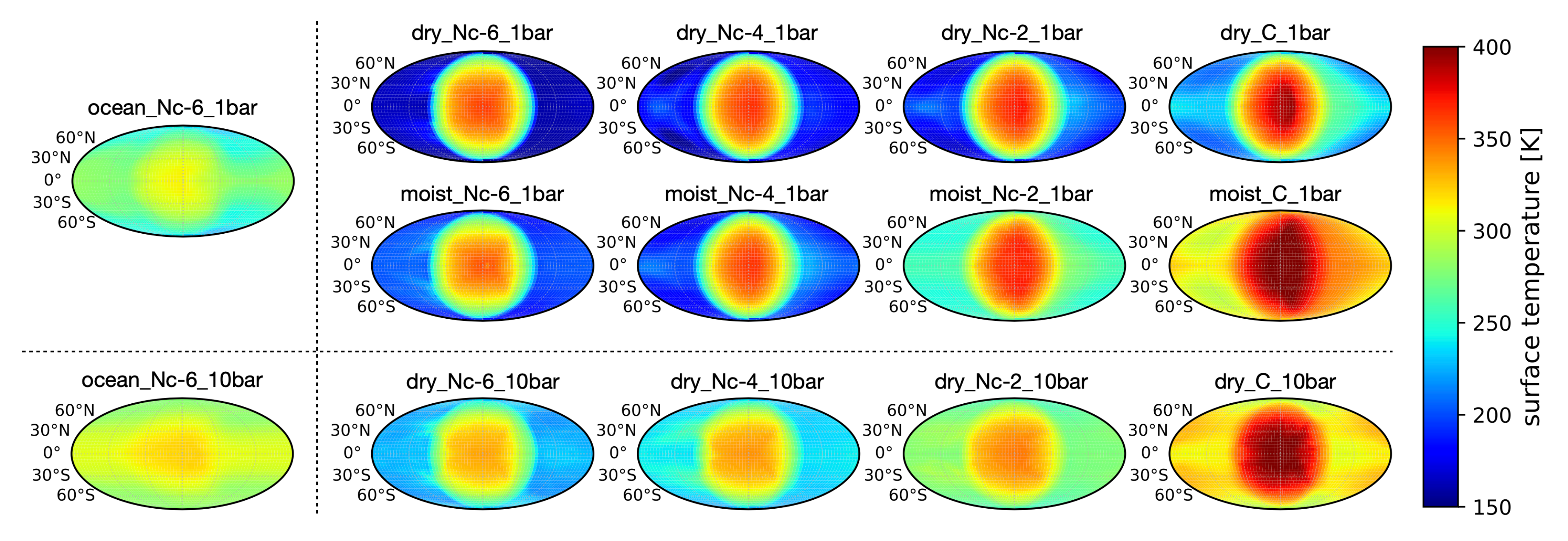}
\caption{Surface temperature maps of our GCM simulations for different surface-atmosphere scenarios (Table~\ref{tbl:models}) with 1~bar (top two rows) and 10~bar (bottom row) atmospheres. The substellar point is located at the center. }
\label{fig:tsurf_1bar}
\end{figure*}

The final step of producing the synthetic observations is to estimate the observational noise. 
To do this, we converted the disk-averaged radiance to the spectral flux observed at a distance of $d=3.83~$pc by multiplying the factor of $\pi R_p^2 /d^2 $, where $R_{\rm p}=1.1R_{\oplus }$ is the planetary radius. 
The spectra are then used as input for the simulator of the observations with \LIFE{}, LIFE\textsc{sim} \citep{LIFE2}, which follows the methodology presented in \cite{LIFE8} and \cite{LIFE12}. 
The model considers relevant astrophysical noises, but does not include instrumental noise effects.
LIFE\textsc{sim} is configured with the parameter values listed in Table \ref{tab:lifesim}, in accordance with previous publications.  
The ratio between the imaging baseline and the nulling baseline is fixed at 6:1, and the length of the nulling baseline is optimized for the target system within the range of 10 and 100~meter such that the modulation efficiency at 10~$\mu $m is maximized for the center of the HZ of each system. 
In the case of Teegarden's Star b, it is adjusted to 100~meter, the upper limit, to best resolve the planet. 

Note that these baseline lengths may be translated to the nominal IWA, $\Delta L$, through $\Delta L \sim \lambda /(2b)$, where $b$ is the length of nulling baseline. 
With the nulling baseline of 100 meters, the IWA would be $\sim 10$~mas at $\lambda = 10$~\micron{}, and $\sim 20$~mas at $\lambda = 20$~\micron{}. 
These values are, in fact, larger than the maximum angular separation, $l_{\rm max}$, of Teegarden's Star b ($l_{\rm max} = 6.8$~mas). 
However, the transmittance of the interferometer interior to the nominal IWA is non-zero, allowing partial signal transmission that can be exploited for detection. 
The angular separation varies with the orbital phase of the planet following $l = l_{\rm max} \sin \alpha $. 
As $\alpha $ deviates from $90^{\circ }$, the signal becomes increasingly affected by noise. This effect is included in our simulations. 

\subsection{Observation strategies}
\label{ss:strategies}

We consider the following two approaches to characterizing thermal emission with respect to the influence of 3D atmospheric structures.

We first examine phase variations, i.e., the dependence of the disk-averaged thermal emission spectra on orbital longitude ($\phi $), or equivalently, on phase angle ($\alpha $). 
Such variations are expected to be among the most direct manifestations of horizontal temperature gradients on exoplanets. 
We quantify their detectability assuming an orbital inclination of $i = 60^{\circ}$, which corresponds to the median value under an isotropic inclination distribution. 
Note that the amplitude at other inclinations can be approximated by scaling with $\sin i$. 

However, phase variations may not always be detectable, as detecting phase variation requires two conditions:
(1) the orbital inclination must not be too low, and (2) the absolute flux of the planetary thermal emission must be accurately measured at each orbital phase. 
In particular, if the orbit is nearly face-on and the obliquity is nearly zero, phase variations will be undetectable.

In light of these limitations, we consider a second approach to characterization: analyzing the information contained in single-epoch (snapshot) thermal emission spectra.
Here, we assume $i=0^{\circ }$ and investigate the relation between the detailed spectral shape of the emission and the 3D atmospheric structure. 
To clarify the 3D effects different from the 1D models, we additionally employ a 1D model \texttt{HELIOS} \citep{Malik+2017,Malik+2019a,Malik+2019b,Whittaker+2022}, assuming the same atmospheric composition and planetary parameters. 

\section{Results for Teegarden's Star b}
\label{s:result}

\subsection{Temperature structures}
\label{ss:result_climate}

Figure \ref{fig:tsurf_1bar} presents the surface temperature maps with different scenarios listed in Table \ref{tbl:models}. 
The equatorial cross sections are presented in Appendix \ref{ap:tsurf_1D} for a more quantitative view of the structure. 
As expected, the horizontal temperature gradient is substantially larger when the planet lacks a global ocean. 
Among dry and moist runs, increasing the abundance of greenhouse gases and/or increasing the atmospheric pressure decreases the temperature contrast between the dayside and nightside (as measured by $(T_{\rm substellar} - T_{\rm antistellar})/T_{\rm ave}$, where $T_{\rm substellar/antistellar}$ is the temperature at the substellar/antistellar point, and $T_{\rm ave}$ is the globally averaged temperature), consistent with \citet{KollAbbot2015} and \citet{WangYang2022}. 
Nevertheless, even for the 100\% CO$_2$ scenarios, the day-night contrast remains substantial relative to the ocean scenarios. 

In the moist runs, the globally averaged water vapor mass mixing ratios just above the surface were found to be  $\sim 5\times 10^{-6}$, $\sim 5\times 10^{-6}$, $\sim 7\times 10^{-4}$, and $\sim 7\times 10^{-3}$, for \texttt{moist\_Nc-6\_1bar}, \texttt{moist\_Nc-4\_1bar}, \texttt{moist\_Nc-2\_1bar}, and \texttt{moist\_C\_1bar} runs, respectively. 
The former two with CO$_2$-poor atmospheres were not efficient in transporting the heat globally, and the atmospheric water vapor was trapped in the cold nightside polar regions. 
As a result, although initialized with abundant atmospheric water vapor, the CO$_2$-poor scenarios ended up with the climate states very similar to the counterpart dry scenarios. 
On the other hand, the latter two with more abundant CO$_2$ retain moist atmospheres, which contribute to the greenhouse effect of the atmosphere and the heat transport. 

Note that the freeze-out of CO$_2$ does not occur in the current parameter space; the sublimation temperature of CO$_2$ at $10^{-6}$, $10^{-4}$, $10^{-2}$, and $1$ bars are 105~K, 125~K, 150~K, and 195~K, respectively \citep{FraySchmitt2009}, which are lower than the minimum temperatures of the corresponding runs. 

\begin{figure*}
\centering
\includegraphics[width=\hsize]{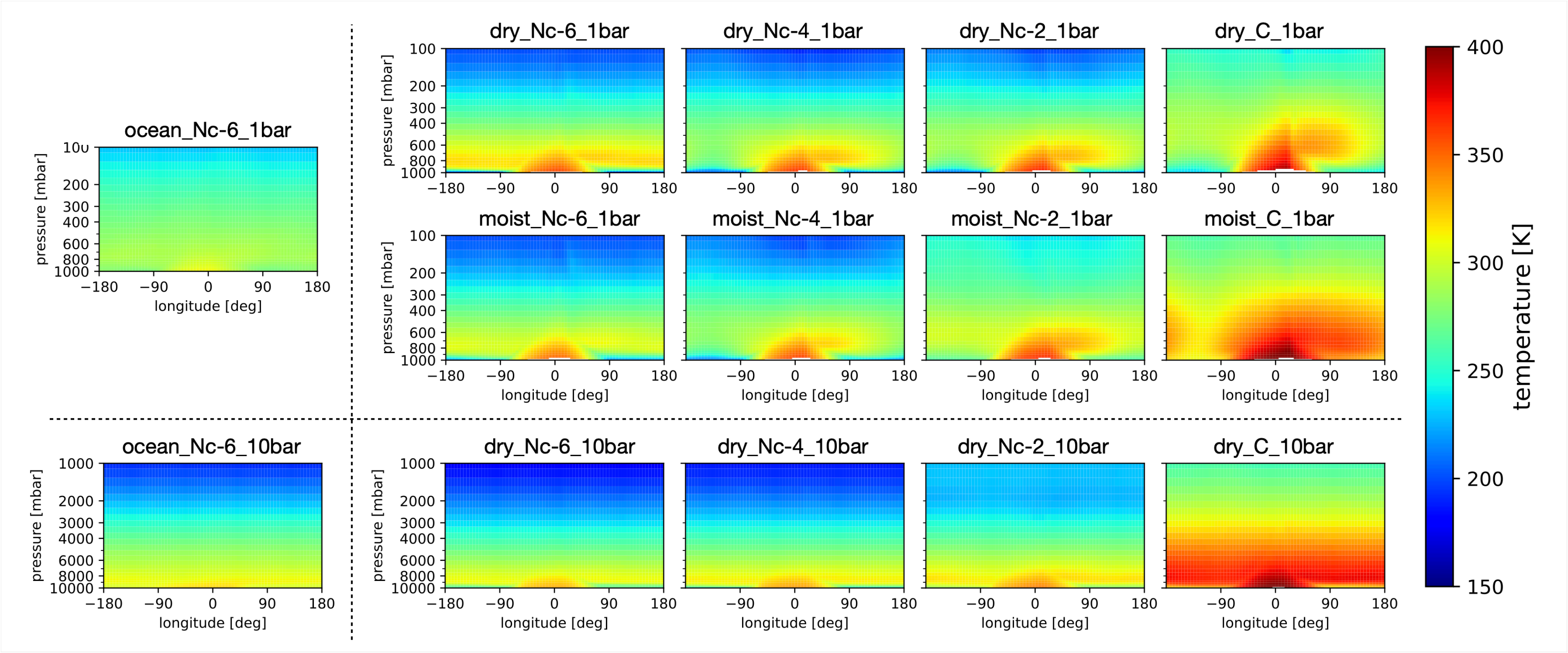}
\caption{Longitudinal–vertical temperature structures in the equatorial region of our GCM simulations for different surface-atmosphere scenarios (Table~1) with 1~bar (top two rows) and 10~bar (bottom row) atmospheres. 
Temperatures are averaged over latitudes within $\pm 10^{\circ}$. 
Longitude $0^{\circ}$ corresponds to the permanent substellar point.}
\label{fig:tb_lon2plm} 
\end{figure*}

Figure \ref{fig:tb_lon2plm} presents the longitudinal–vertical temperature structure in the equatorial region for all scenarios, averaged over latitudes within $\pm 10^{\circ}$. 
They commonly show the thermal inversion layer near the bottom of the atmosphere in regions away from the substellar point. 
Such a thermal inversion away from the subsellar point is a well-known characteristic of tidally locked planets \citep{Joshi+1997}. 
The temperature inversion is more prominent in the dry runs, associated with the large temperature gradient across the surface. 

\begin{figure}
\centering
\includegraphics[width=\hsize]{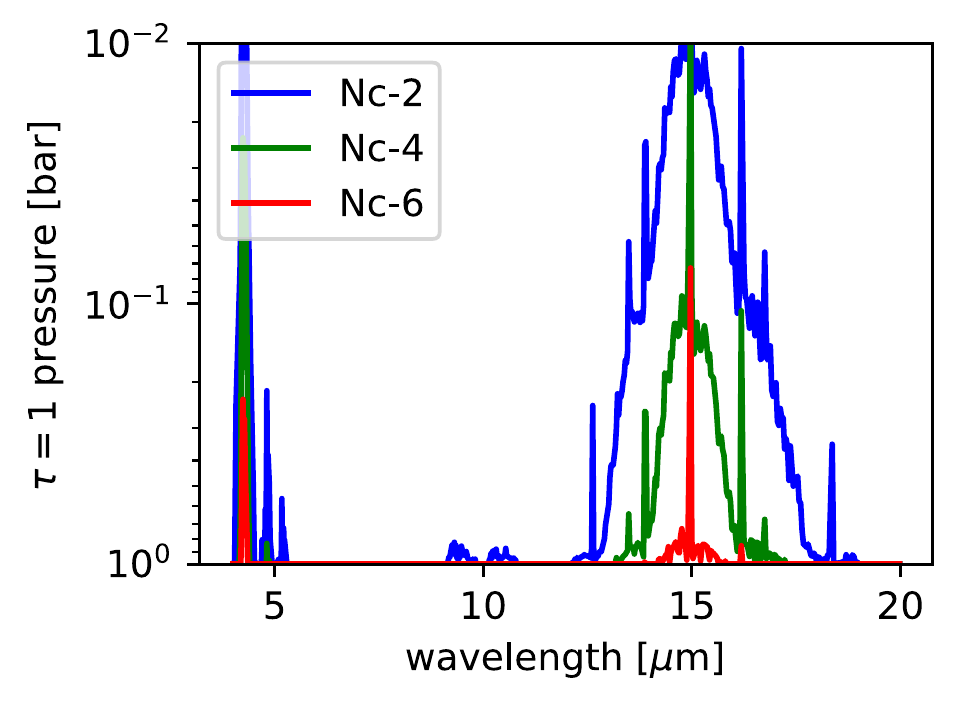}
\caption{The pressure levels where the vertical optical depth from the top of atmosphere becomes unity.}
\label{fig:photosphere} 
\end{figure}

\begin{figure}[h!]
\centering
\includegraphics[width=\hsize]{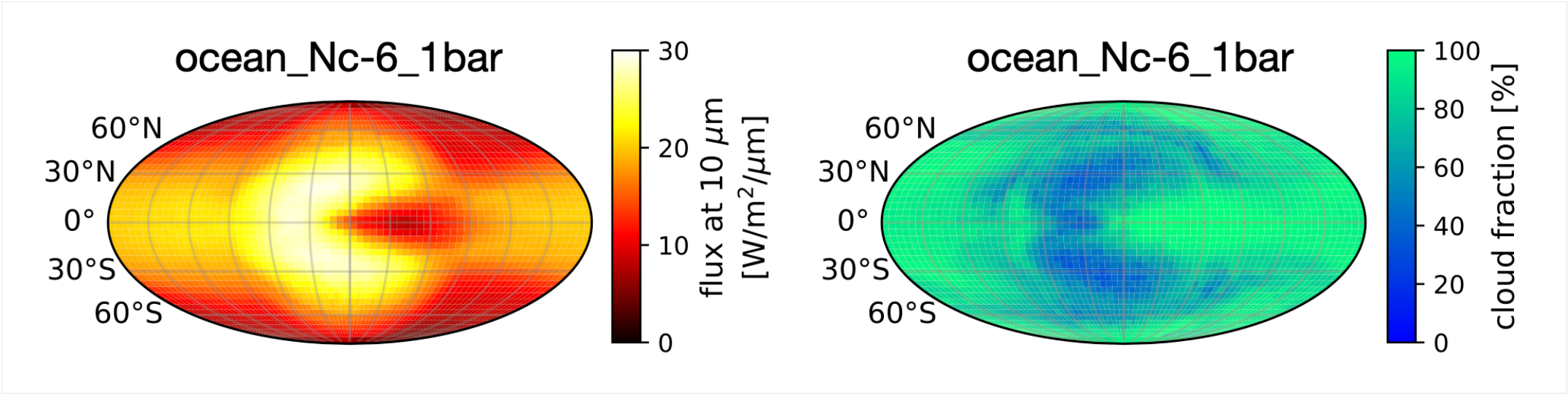}
\caption{Outgoing flux at $\lambda= 10\,\mu $m (left) and the total cloud cover fraction (right) in the case of \texttt{ocean\_Nc-6\_1bar} run. The substellar point is located at the center. }
\label{fig:cldt_H_1000mb}
\end{figure}

The outgoing radiation follows these temperature patterns along the wavelength-dependent photosphere; the latter is shown in Figure \ref{fig:photosphere} for the clear-atmosphere cases. 
For ocean and moist scenarios, the outgoing radiations are modified by cloud decks. 
This can be confirmed by comparing the surface temperature map, the map of 10~\micron{} outgoing radiation, and the map of cloud cover—the latter two are shown in Figure~\ref{fig:cldt_H_1000mb}—where a strong correlation is seen between outgoing flux and cloud distribution. 

The cloud pattern for the ocean planet (\texttt{ocean\_Nc-6}) is qualitatively consistent with climate simulations with other GCMs of similar generation, and with recent simulations with cloud-resolving models \citep{Yang+2023}. 
Specifically, while clouds in general develop near the subsetellar point where the surface temperature is highest, 
the cloud morphology on habitable planets orbiting late M-type stars is influenced by planetary rotation
\citep{Kopparapu+2017,Haqq-Misra+2018}. 
Due to the development of zonal wind as well as weakened substellar convection, the cloud decks in the substellar regions are elongated toward the East, 
shifting the region of highest cloud optical thickness away from the hottest point (see the right panel of Figure \ref{fig:cldt_H_1000mb}). 

\subsection{Phase variations of thermal emission spectra}
\label{ss:result_phasecurve}

\begin{figure*}
\centering
\includegraphics[width=\hsize]{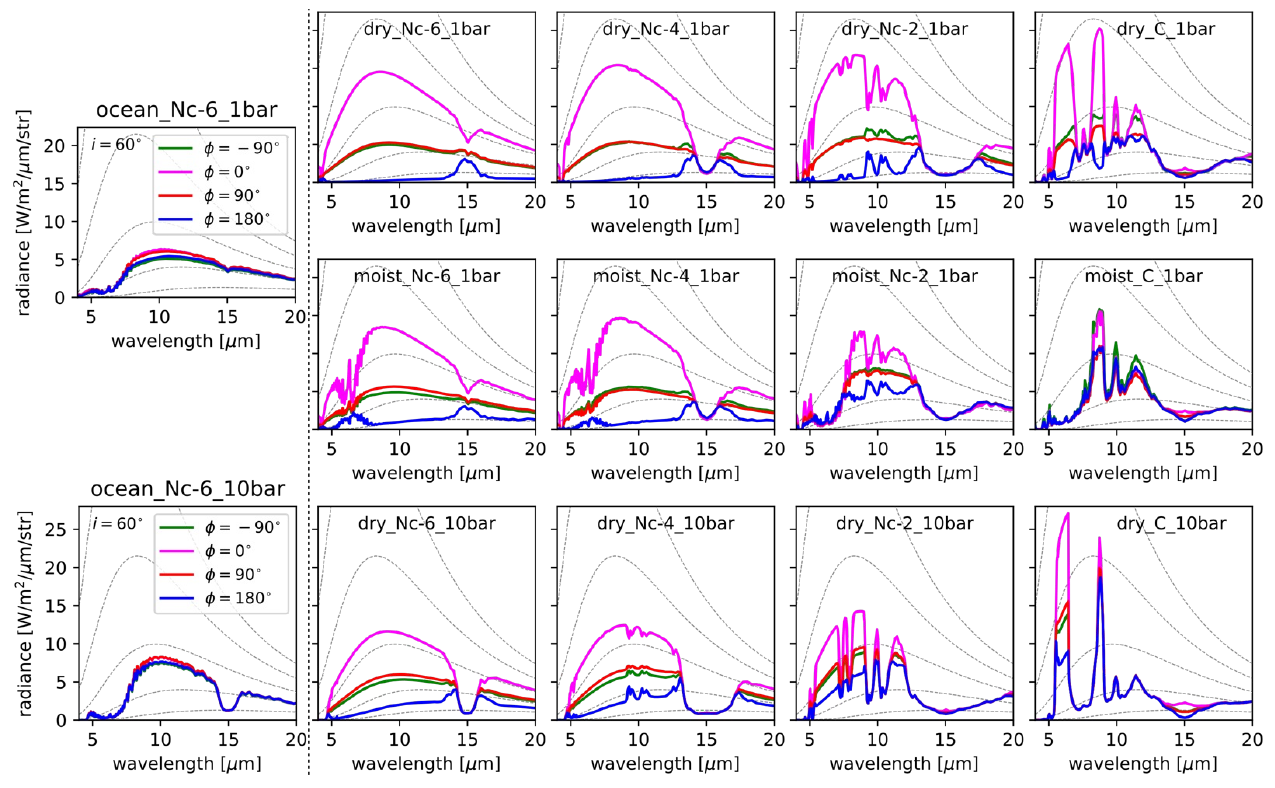}
\caption{Thermal emission spectra of all scenarios considered in this study at four orbital longitudes ($\phi =-90^{\circ},\,0^{\circ},\,90^{\circ},\,180^{\circ}$) with an orbital inclination of $i=60^{\circ }$. 
Here, the orbital longitude ($\phi$) is measured from the point of maximum dayside visibility; in the case of an edge-on orbit, $\phi =0^{\circ}$ and $\phi=180^{\circ}$ correspond to the eclipse and the transit, respectively, and $\phi = \pm 90^{\circ}$  correspond to the observations from above the eastern and western terminators relative to the substellar point. 
The dotted lines correspond to the black body emission at 200~K, 250~K, 300~K, 350~K, and 400~K from bottom to top. Note the different range of $y$-axis between the upper two and lower panels. }
\label{fig:phase_variation}
\end{figure*}

\begin{figure}
\centering
\includegraphics[width=\hsize]{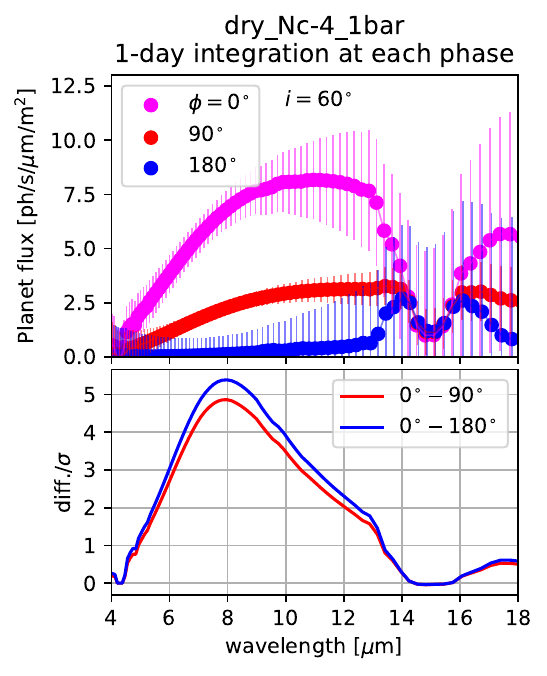}
\caption{The upper panel presents the signals and errors from mock observations for the \texttt{Nc-4\_1bar} scenario of Teegarden's Star b at different orbital longitudes ($\phi$) assuming a 1-day integration at each orbital phase and an orbital inclination of $i=60^{\circ}$. 
The lower panel shows the flux differences between two orbital longitudes normalized by the corresponding errors. 
The orbital longitude ($\phi$) is measured from the point of maximum dayside visibility; for an edge-on orbit, $\phi =0^{\circ}$ and $\phi=180^{\circ}$ correspond to the eclipse and the transit, respectively.}
\label{fig:Nc-4_SNR}
\end{figure}

\subsubsection{Thermal emission spectra and  absorption bands}

Figure \ref{fig:phase_variation} presents the disk-averaged thermal emission spectra of all the scenarios at four different orbital longitudes: $\phi =-90^{\circ}, 0^{\circ },\,90^{\circ },$ and $180^{\circ }$, assuming an orbital inclination of $i=60^{\circ }$. 
Thus, these orbital longitudes correspond to phase angles of $\alpha =90^{\circ },\,30^{\circ },\,90^{\circ },\,150^{\circ }$.

Before addressing phase variations, we first describe the characteristics of the molecular absorption bands in these spectra. 
As the CO$_2$ mixing ratio increases (from left to right in each row), the 15~$\mu$m CO$_2$ absorption band becomes deeper, and double-peaked absorption features centered at 10~$\mu$m emerge when the partial pressure of CO$_2$ exceeds $\sim 10^{-3}$~bar.
When the partial pressure of CO$_2$ exceeds $10^{-2}$~bar,  CO$_2$-CO$_2$ Collision Induced Absorption (CIA) features around 8~$\mu $m also start to appear. 
Due to the enhanced greenhouse effect, models with higher CO$_2$ mixing ratios exhibit a higher continuum. 

The absorption feature at 6~\micron{} is due to water vapor, and is commonly seen in ocean and moist no-ocean cases, with varying strengths. 
This implies that the presence of the 6~\micron{} H$_2$O band alone does not necessarily indicate the existence of a surface ocean. 
Meanwhile, the strength of water vapor features are closely related to the surface temperature gradient. 
Planets that develop large temperature contrast on the surface, such as the \texttt{moist\_Nc-6\_1bar} and \texttt{moist\_Nc-4\_1bar} runs, cannot sustain an water vapor partial pressure of $\sim 10^{-2}$~bar. 
In turn, strong water vapor absorption features imply the efficient heat re-distirbution.

\subsubsection{Large phase variation as anti-indicator of a global ocean}

As shown in Figure \ref{fig:phase_variation},  the thermal emission of the ocean-covered scenarios has minimal dependence on the orbital phase. 
In contrast, the no-ocean scenarios exhibit significant variations in the continuum level outside the absorption bands where emission originates from surface, reflecting the large temperature contrast of the surface (Figure \ref{fig:tsurf_1bar}). 
The suppressed phase variations within the absorption bands are due to the fact that these wavelengths probe high-altitude atmospheric layers where temperatures are more uniform \citep[e.g.,][see also Figure~\ref{fig:tb_lon2plm}]{Selsis+2011} . 
The exact variation amplitude outside the molecular absorption bands depends on the atmospheric scenarios and wavelength. While the temperature contrast monotonically decreases as the greenhouse effect of the atmosphere increases, the nonlinear dependence of thermal emission on the temperature can cause large variation in the Wien regime with moderate temperature contrast. 
Notably, even in the 1-bar CO$_2$-dominated case, where the day–night temperature contrast is relatively modest 
(the day-night temperature difference is $\sim 100$~K, compared to $\gtrsim 200$~K in other 1-bar cases), significant variation is still seen at $< 10$~$\mu $m. 
For the 10-bar CO$_2$ case, variability is further suppressed at 8–9~$\mu$m, and flux variations at even shorter wavelengths (5–7~$\mu$m) remain important for constraining surface temperature gradients.  

In moist runs, the wavelength range over which the flux varies is limited by the strong water vapor absorption band at 6 \micron{}. In addition, the maximum flux variations are smaller than those of the dry counterparts, consistent with the surface temperature maps. 
However, they are still larger than the global-ocean scenarios within the parameter space considered. 

\subsubsection{Detectability}

We now quantify the detectability of phase variations. 
The upper panel of Figure~\ref{fig:Nc-4_SNR} compares the signal and noise for mock observations of Teegarden’s Star b under the \texttt{dry\_Nc-4} scenario at four orbital phases ($\phi = \{-90^{\circ },\,0^{\circ },\,90^{\circ },\,180^{\circ }\}$, or $\alpha = \{ 90^{\circ },\,30^{\circ },\,90^{\circ },\,150^{\circ }\}$), assuming 1-day integrations at each orbital phase. 
Given the planet’s orbital period of approximately 4.9 days, some orbital motion occurs during each integration, though this effect is not accounted for in the current simulation. 
It can be recognized that the error bars at $\phi = 0^{\circ}$ and $180^{\circ}$ ($\alpha = 30^{\circ }$ and $150^{\circ }$) are larger than those at $\phi = 90^{\circ}$ ($\alpha = 90^{\circ}$) despite identical integration times. 
This difference arises from the variation in angular separation from the host star: $\phi =\,(\alpha =)\, 90^{\circ}$  corresponds to the maximum elongation and therefore experiences the least stellar contamination because of better transmission map. 
The lower panel of Figure \ref{fig:Nc-4_SNR} shows the difference between two orbital phases normalized by the errors. 
For the \texttt{dry\_Nc-4\_1bar} scenario, the spectra at different orbital phases differ by up to $\sim $4~$\sigma$, and identifying such variations is straightforward. 
In this case, the SNRs peak near $\lambda = 8$~\micron{}. 
While the precise location of the peak depends on the specific planetary properties and the configuration of the interferometer, the general trend is expected to hold: at shorter wavelengths, the SNR tends to be limited by the intrinsically weak planetary signal (and potential stellar leakage), whereas at longer wavelengths, the SNR is typically degraded by increased contamination from exozodiacal emission and the larger IWA (i.e., reduced transmission of the planetary signal). 

\begin{figure}[hbt]
\centering
\includegraphics[width=\hsize]{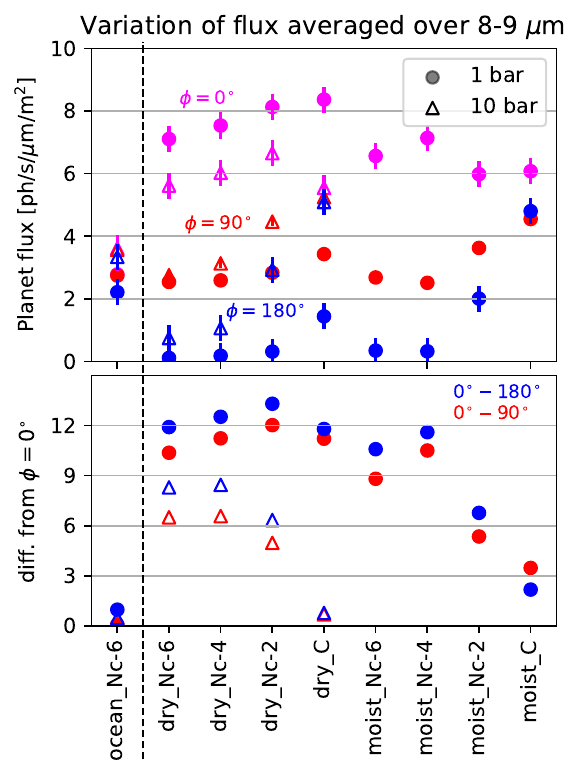}
\caption{The upper panel presents planetary fluxes averaged over 8-9 $\mu $m for different surface scenarios at various orbital longitudes, assuming an orbital inclination of $i=60^{\circ }$. 
The lower panel shows the flux differences between $\phi = 0^{\circ }$ and $180^{\circ }$ (blue) or $\phi = 90^{\circ }$ (red). 
The orbital longitude ($\phi$) is measured from the point of maximum dayside visibility; for an edge-on orbit, $\phi = 0^{\circ}$ and $\phi = 180^{\circ}$ correspond to the eclipse and the transit, respectively. }
\label{fig:varamp_detectability_TGb}
\end{figure}

In further evaluating the detectability of phase variations for other scenarios, we leverage the fact that detecting phase variations does not require high spectral resolution; the flux can be averaged over selected bandpasses to improve SNR. 
To optimize photon collection while avoiding overlap with strong molecular absorption features, we propose using the variation of flux averaged over the 8–9~\micron{} as the primary metric. 
The upper panel of Figure~\ref{fig:varamp_detectability_TGb} presents the band-averaged radiance over 8–9~$\mu$m for different scenarios at $\phi = 0^{\circ},\,90^{\circ},$ and $180^{\circ}$, with error bars corresponding to 1-day integrations, as in Figure~\ref{fig:Nc-4_SNR}. 
In the lower panel, the blue (red) points represent the difference between $\phi = 0^{\circ }$ and $180^{\circ }$ ($\phi = 0^{\circ }$ and $90^{\circ }$), normalized by the corresponding errors. 
It can be seen that in most cases the difference between $\phi = 180^{\circ }$ and $0^{\circ }$ is slightly more advantageous in terms of SNR, despite the larger intrinsic noise caused by smaller angular separations from the host star; these cases will be referred to in the following.  
For dry runs, the flux values at $\phi = 180^{\circ }$ and $0^{\circ }$ are separated by more than $5\sigma$ except for \texttt{dry\_C\_10bar}. 
The flux variations of the 1-bar moist runs tend to be smaller than dry counterparts, but remain recognizable by $>5\sigma $ for cases with minor CO$_2$, and by $3\sigma $ for the CO$_2$ 100\% case. 
Additionally, we also find that the use of the flux variation in the 5.5-6.5~\micron{} range as the secondary metric successfully identifies the flux variation for the \texttt{dry\_C\_10bar} case by  $\sim 9\sigma$. 
These results indicate that multi-epoch observations with 1-day integration at each orbital phase will be able to detect flux variations for the majority of $\lesssim 10$~bar dry planets or $\lesssim 1$~bar moist scenarios. 

\subsection{Spectral shape of snapshot thermal emission spectra}
\label{ss:result_snapshot}

\subsubsection{3D effects on snapshot spectra as anti-indicator of ocean}

Considering cases where phase variations cannot be well constrained, this section explores the clues in snapshot thermal emission spectra. 
As an illustration of such cases, here we assume a face-on planetary orbit ($i=0^{\circ}$). 

The colored lines of Figure \ref{fig:snapshot_Nc1000mb_1D3D} present the snapshot spectra based on three of our runs with $i=0^{\circ }$. 
The coral one represents the spectrum for \texttt{dry\_Nc-4\_1bar}
which is characterized by a large day-night surface temperature contrast. 
Note that this spectrum closely resembles the green line in the \texttt{dry\_Nc-4\_1bar} panel of Figure~\ref{fig:phase_variation}, with the only difference being the orbital inclination. 
For comparison, the spectrum calculated with the 1D model \texttt{HELIOS}, using the same atmospheric and planetary parameters, is shown by the gray line. 

The comparison between the spectra based on 3D and 1D models reveals two key differences.  
First, the continuum shape based on the 3D model deviates from that of a blackbody. 
Specifically, there is an excess at wavelengths shorter than the peak flux ($\sim 10$~\micron{}), and a deficit at longer wavelengths. 
This appears as an overall decrease in brightness temperature (lower panel of Figure \ref{fig:snapshot_Nc1000mb_1D3D}) with increasing wavelength, except within the absorption bands. 
This feature arises from the horizontal temperature gradient, and is not seen in the spectra based on the 1D model. 
A similar trend has been observed in the disk-averaged thermal emission spectra of Mars, which has a tenuous atmosphere \citep{Lellouch+2000}, and more recently in the hot Jupiter WASP-69~b \citep{Schlawin+2024}. 
This effect is further illustrated using a toy model in Appendix~\ref{ap:2Dtoymodel}.  
The overall decreasing brightness temperature of the continuum is indicative of a horizontal temperature gradient, which would not be compatible with the presence of a global ocean. 

One can also observe a similar decrease in brightness temperature in the ocean-covered case (\texttt{ocean\_Nc-6\_1bar}; sky blue), albeit to a lesser degree. 
This is mainly due to H$_2$O absorption, which becomes more pronounced at longer wavelengths, rather than a surface temperature gradient. 
Because of increased atmospheric opacity, the photosphere shifts to higher--and thus cooler--altitudes at longer wavelengths, resulting in a decrease in brightness temperature. 
This introduces a new type of false positive for identifying H$_2$O absorption bands at wavelengths beyond 10~$\mu$m, which will be discussed further in Section~\ref{ss:dis_Bias1D}. 

The second difference between the 3D and 1D models is the shape of absorption features, particularly near the edges of the absorption bands.
Specifically, there are bumps at $\sim$14~$\mu$m and $\sim$16~$\mu$m in the 3D spectrum (\texttt{Nc-4}) in Figure~\ref{fig:snapshot_Nc1000mb_1D3D}.
These arise from the combination of the wavelength-dependent photospheric depth (Figure~\ref{fig:photosphere}) and the thermal inversion near the surface away from the substellar point (see Section~\ref{ss:result_climate} and the left two panels of Figure~\ref{fig:tb_lon2plm}).
More specifically, this peculiar shape of the ``absorption'' band can be interpreted as follows:
At the center of the absorption bands, where the atmosphere is optically thick, the emission originates from cool high altitudes.
In the continuum, where the atmosphere is transparent, emission arises from the surface, which is hot near the substellar point but cold elsewhere.
At wavelengths where the atmosphere is slightly opaque, emission originates from altitudes just above the surface.
These layers are cooler than the dayside surface but warmer than the nightside surface, and the average emission may exceed that from the surface itself, producing surges at the edges of absorption bands.
Note that the surges become more prominent as the nightside becomes more visible, as illustrated by the spectra for $i=60^{\circ}$ and $\phi=180^{\circ}$ (blue lines in Figure \ref{fig:phase_variation}). 
While these features do not directly constrain horizontal temperature gradients, they indicate the thermal inversion layer just above the surface which is stronger as the surface temperature contrast increases, and suggest that the observer is seeing down to the surface of a synchronously rotating planet. 

\subsubsection{Detectability}

\begin{figure}[h]
\centering
\includegraphics[width=\hsize]{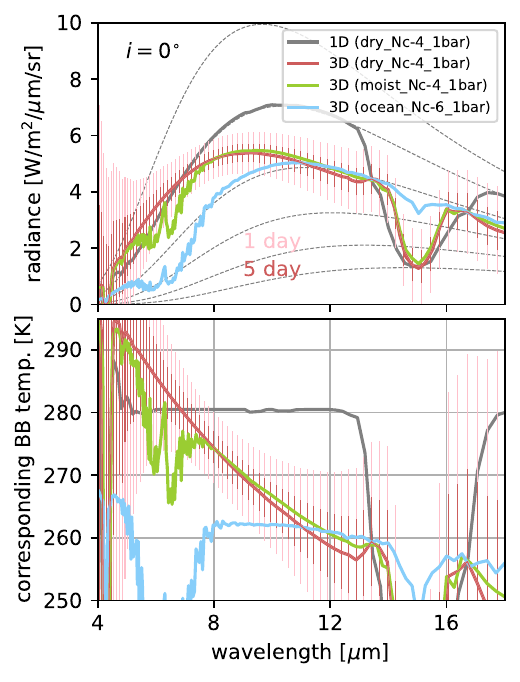}
\caption{The upper panel compares the modeled disk-averaged thermal emission spectra under various assumptions, together with the errors from 1-day and 5-day observations.  The coral, yellow-green, and light-blue lines correspond to
\texttt{dry\_Nc-4\_1bar}, \texttt{moist\_Nc-4\_1bar} and \texttt{ocean\_Nc-6\_1bar}, respectively, with $i=0^{\circ }$ (face-on orbit). 
The gray line is based on the 1D temperature-pressure profile in radiative-convective equilibrium calculated with HELIOS for the same atmospheric composition as \texttt{dry\_Nc-4\_1bar}. 
The black dotted lines represent blackbody emission at 200~K , 220~K, 240~K, 260~K, 280~K, and 300~K from bottom to top. 
As the coral and gray lines assume the same atmospheric composition and irradiation, the difference between the two illustrates the effect of the 3D temperature structure. 
The lower panel shows the same set of spectra expressed in terms of the brightness temperature at each wavelength.}
\label{fig:snapshot_Nc1000mb_1D3D}
\end{figure}

The error bars in Figure~\ref{fig:snapshot_Nc1000mb_1D3D} are based on LIFE\textsc{sim} assuming 1 day (light pink) or 5 days (coral) of integration. 
An implicit assumption here is that the spectrum does not change during the integration time, as expected for synchronously rotating planets on face-on orbits. 
The required integration times considered are too long to be considered ``snapshot'' spectra; the term ``snapshot'' is used simply to refer to the key features contained in the spectral shape at a given point in the orbit.

To quantify the detectability the changing brightness temperature of the continuum, we calculated a metric defined by
\begin{equation}
    \hat{\chi}^2 = \frac{1}{J-1} \sum _{\lambda _{\rm min} < \lambda _j < \lambda_{\rm max}} \frac{[ T_b(\lambda _j) - \bar{T_b} ]^2}{\sigma^2 (\lambda _j )} , \label{eq:chi2}
\end{equation}
where $\lambda _{\rm min}$ and $\lambda _{\rm max}$ controls the wavelength range we focus on, $J$ is the number of data points within the range, $T_b(\lambda _j)$ represents the brightness temperature at the wavelength $\lambda _j$, $\bar{T_b}$ is their average (i.e., $\bar{T_b} = (1/J)\sum _{\lambda _{\rm min} < \lambda _j < \lambda_{\rm max}} T_b(\lambda _j) $), and $\sigma _j$ indicates the associated errors. 
The $\hat{\chi }^2$ would correspond to the conventional reduced chi-square metric for the flat (constant-$T_b$) model when evaluated using mock data of $T_b(\lambda _j)$. 
Here, instead, we substitute the noise-free spectral brightness temperatures based on the 3D model for $T_b(\lambda _j)$ and use the resulting $\hat{\chi }^2$ as an indicator metric of whether deviations from a constant-$T_b$ model would be detectable given the assumed noise level. 
Using the noise properties of a the 3-day observation between 5 and 12~\micron{}, we obtained $\hat{\chi^2}\sim 3$
with the degree of freedom of 43. 
Under the null hypothesis of a constant $T_b$, 
this would correspond to a p-value $\ll 10^{-8}$,
roughly equivalent to a $>5\sigma$ one-sided Gaussian significance. Such a result would indicate that a constant-$T_b$ model would be rejected. 
Thus, we estimate that a total integration time 
longer than that required to detect phase variation is needed to obtain the first clues in snapshot spectra. 
Characterization of the shape of absorption bands (i.e., 15~\micron{} CO$_2$ band) likely requires even higher precision as demonstrated in Figure \ref{fig:snapshot_Nc1000mb_1D3D}. 

Note that if a narrower bandpass can be used due to the atmospheric absorption of the target (or the atmospheric windows in the case of ground-based observations), the required precisions become higher. 
For example, if we focus on the bandpass between 8-12 \micron{} (8-9~\micron{}) with the same spectral resolution, the indication of the deviation from the constant-$T$ assumption by 3$\sigma $ would require $\sim $5 ($\sim $60) days of integration. 

\section{Application to Other Systems: Estimated number of candidates}
\label{s:observability}

In this section, we discuss the detectability of the above-mentioned effects of horizontal temperature gradient for other planetary systems.

\begin{figure}
\centering
\includegraphics[width=\hsize]{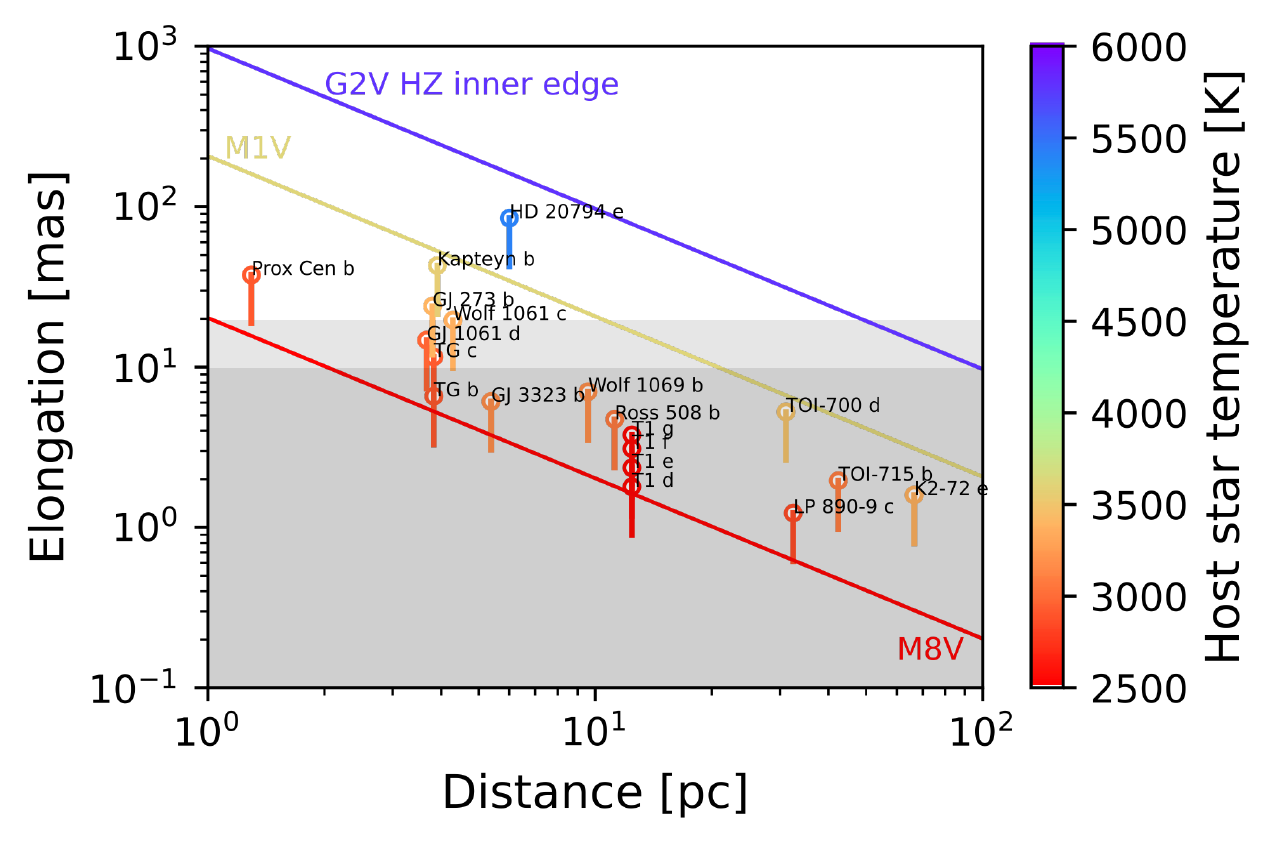}
\caption{Relation between the elongation and the distance from Earth for the inner edge of the habitable zones of G2- (blue), M1- (gold) and M8-type (red) stars. The stellar radius, $R_{\star }$, and the effective temperature, $T_{\rm eff}$, are assumed to be $\{ R_{\star },\,T_{\star,\,{\rm eff}}\} = \{ R_{\odot },\,5800~{\rm K}\}$, $\{ 0.5R_{\odot },\,3600~{\rm K}\}$, $\{ 0.1R_{\odot },\,2500~{\rm K}\}$ for an G2-, M1-, and M8-type star, respectively. The inner edge here is calculated based on \citet{Kopparapu+2013}. 
Known planets are overlaid with labels and color-coded according to the effective temperature of their host stars. 
The vertical lines indicate the range of elongation as the planets move between $30^{\circ }<\alpha < 150^{\circ}$. The gray filled region indicates that the elongation is smaller than the IWA, assumed to be 10~mas at $\lambda = 10$ \micron{} and 20~mas at $\lambda = 20$ \micron{}. }
\label{fig:elong_distance}
\end{figure}

Figure \ref{fig:elong_distance} shows the discovered temperate planets in the solar neighborhood as a function of separation from the host star (elongation) and the distance from Earth, color-coded by stellar effective temperature. 
The planet data are based on the Open Exoplanet Catalogue\footnote{\url{https://www.openexoplanetcatalogue.com/}} as of June 2025. 
The targets are selected based on the mass/radius and the orbital distance: $M_{\rm p}\sin i <5M_{\oplus }$ or $M_{\rm p}<1.5R_{\oplus }$ ($R_{\rm p}$: planetary radius, $M_{\rm p}$: planetary mass), and 
the semi-major axis being larger than 80\% of the inner edge of habitable zones calculated based on 1D models \citep{Kopparapu+2013}. 
The vertical line drawn from each planet indicates the range of elongation corresponding to the range of phase angle between $30^{\circ }$ and $150^{\circ }$, equivalent to the range studied in Section \ref{ss:result_phasecurve}. 
There are approximately 10 known planets around M-type stars within 10 parsecs, with the elongation comparable to or larger than that of Teegarden's Star b. 
They would be the immediate candidate targets of \LIFE{}.

There are also yet-to-be-discovered potentially habitable planets in the solar neighborhood. 
Based on the number or M-type stars within 10 parsecs, 280 \citep{HenryJao2024}, and the occurrence rate of potentially habitable planets around M-type stars, $\sim 0.2$ \citep[e.g.,][]{DressingCharbonneau2015}, the expected number of potentially habitable planets within 10 (5) parsecs is $\sim $56 ($\sim $7). 
This suggests that a few more targets could be discovered within 5~parsecs, while a larger discovery space exists between 5 and 10 parsecs. 

A requirement specific to detecting phase variation is the need for observations at multiple orbital phases with significantly different phase angles. 
This means that only those with inclined orbits are practical targets for phase variation studies. 
Using Teegarden's Star b as a reference case, we adopt $i > 60^{\circ}$ as a sufficient inclination for detecting variations at a similar level. 
Statistically, assuming random orbital orientations, the probability of a planet having $i > 60^{\circ}$ is 50\%. 
Thus, approximately half of the nominal \LIFE{} targets should satisfy this criterion, leaving a few to several promising targets within 5~parsecs.  

In addition to the inclined orbit, the elongation needs to be sufficiently large even at the orbital phase away from quadrature. 
The solid lines in Figure~\ref{fig:elong_distance} show the maximum elongation of planets at the inner edge of the habitable zone for M8- (red), M1- (gold), and G2-type (blue) stars as a function of distance from Earth. 
The inner edge of the habitable zone is calculated using the analytic formula from \citet{Kopparapu+2013}, 
and the stellar radii and effective temperatures are assumed to be $\{ R_{\star},\,T_{\star,\,\rm eff} \} = \{ R_{\odot},\,5800\rm K \}$, $\{ 0.5 R_{\odot},\,3600~\rm K \}$, and $\{ 0.1 R_{\odot},\,2500~\rm K \}$ for G2-, M1-, and M8-type stars, respectively. 
The comparison between these lines and the fiducial IWA of \LIFE{} (shadowed) suggests that the IWA would significantly affect the observations for those around mid- and late-M stars beyond 5~parsecs, and those around early-M stars beyond 10~parsecs. 

Of course, the integration time required to reach the necessary SNR varies from target to target. 
Simulated observations of a planet orbiting an M3-type star at 5 parsecs indicate that approximately 1.5 days of integration yield the data quality equivalent to Figure \ref{fig:varamp_detectability_TGb}.  
Thus, we estimate for planets within $\sim 5$ parsecs that 1-2 days of integration at each orbital phase would identify the band-averaged phase variation of a dry scenario with an atmospheric pressure of $\lesssim 10$~bar, while the targets at 10 parsecs will require about 4 times longer integration time. 
As studied in Section \ref{s:result}, detection of finer 3D effects in snapshot spectra will require integration times a few times longer than those quoted above.

\section{Discussion}
\label{s:discussion}

\subsection{For which planets does temperature gradient work as anti-indicator of ocean?}

\subsubsection{Dependence on incident flux}
\label{sss:dep_Sx}

The degree of enhancement in heat transport efficiency due to the presence of a surface ocean depends critically on the amount of stellar irradiation absorbed by the planet. 
The effect is weaker for cooler planets, owing to the reduced heat transport capacity of water vapor.
This trend is clearly demonstrated in \citet{Yang+2013} and \citet{Wolf+2019} for scenarios in which planets with 1~bar N$_2$-dominated atmospheres with zero-flux oceans orbit M-type stars. 
Indeed, the variation amplitude of our dry/moist runs with 1~ppm CO$_2$ are similar to that of the ``cold'' ($T_{\rm eq}\sim 200$~K) aqua planet runs of \citet{Wolf+2019} (see the left column of their Figure 8). 
Although ocean dynamics can contribute to reducing the day–night temperature contrast \citep{DelGenio+2019}, the general trend that cooler planets can develop relatively larger day–night temperature contrasts likely holds. 
The trend that phase variations of ocean planets are more muted at higher insolations is also expected to apply to condensible-rich atmospheres, such as those with minimal background gases \citep[e.g.,][]{DingPierrehumbert2018,DingPierrehumbert2020}.
Therefore, planets closer the inner edge of the habitable zone offer a more robust opportunity to assess the absence of a global ocean, with Teegarden's Star b being one of the ideal targets. 
The small incident flux of Teegarden's Star c (zero-albedo $T_{\rm eq}\sim 210$~K) would make it challenging to distinguish the scenarios.  

We note, however, that regardless of its magnitude, a planet with a global ocean would exhibit a milder temperature contrast than one without, independent of the incident flux, when other parameters are equal. 
Thus, phase variations or the 3D effects on snapshot spectra remain valuable indicators when evaluating the possibility of a global ocean. 
In general, constraining the presence of a global ocean may be approached by assessing whether a 3D model with or without an ocean better reproduces all observed thermal emission characteristics, including greenhouse gas composition, water vapor abundance, and the spatial gradient of top-of-atmosphere emission, given the boundary conditions such as incident flux and rotation period. 

\subsubsection{Dependence on spectral types of the host star}
\label{sss:dep_SpType}

While we have focused on the synchronously rotating planets around M-type stars,
the general concept that surface temperature gradients imprint characteristic features on thermal emission spectra can in principle be applied to planets around other types of stars as well. 
Indeed, both Mars and the Moon, which have little or no atmosphere, develop significant surface temperature gradients and exhibit a decrease in brightness temperature in the mid-infrared domain toward longer wavelengths \citep[e.g.,][]{Lellouch+2000, Muller+2020}, along with large phase variations \citep[e.g.,][]{Tinetti+2005,Robinson2011}. 

An advantage of planets orbiting low-mass stars is that, as noted in Section~\ref{ss:method_target}, their rotation periods are theoretically better constrained due to the shorter tidal-locking timescales. 
In particular, the high likelihood of synchronous rotation facilitates the interpretation of their thermal emission properties. 
Theoretical estimates predict that habitable-zone planets around M- and K-type stars can be synchronized within a few Gyr in the absence of satellites \citep{Barnes+2017}. 
While planets around M-type stars are likely more abundant, the larger elongation of habitable-zone planets around K-type stars from the host star (Figure \ref{fig:elong_distance}) also makes them promising targets for \LIFE{}. 
This view aligns with the concept of the ``K-dwarf advantage'' \citep{HellerArmstrong2014, Arney2019, LoboShields2024}.

We note that the climate regime of synchronously rotating planets varies with the rotation period. 
The rotation periods of synchronously rotating habitable-zone planets range from several days to more than 100 days for systems around M9- to K-type stars. Our benchmark case, with a rotation period of $\sim 5$ days, corresponds to near the shortest end of this range and develops banded zonal winds that displace the substellar cloud deck eastward, consistent with the atmospheric structures seen in other GCM studies of ``rapid rotators'' \citep{Yang+2014, Kopparapu+2017, Haqq-Misra+2018, Wolf+2019}.
As the rotation rate decreases, the zonal winds extend toward the poles, the surface temperature distribution becomes more symmetric about the substellar–antistellar axis, and thicker cloud decks form near the substellar point. \citet{Haqq-Misra+2018} showed that the resulting day–night surface temperature contrast of synchronously rotating planets orbiting early- to mid-M-type stars (i.e., ``slow rotators'') are comparable to that of planets around late-M-type stars near the inner edge of the habitable zone. Owing to the thick substellar cloud decks, the thermal phase curves of slow rotators near the inner edge of the habitable zone can even be slightly inverted, that is, the disk-integrated thermal emission flux reaches its minimum when the observer faces the substellar point \citep[e.g.,][]{Yang+2013, Haqq-Misra+2018,Wolf+2019}. 
The effects of the rotation period on dry and moist scenarios are limited, and they develop a large day–night temperature contrast in the slow-rotation regime as well \citep[e.g.,][]{Yang+2013,Lobo+2023}. 
Nearly flat or inverted thermal phase curves of ocean-covered slow rotators would therefore be distinguishable from ocean-free scenarios, in a manner similar to our case study.

The rotation periods of planets orbiting G-type stars are more uncertain, making signal interpretation more challenging. 
Qualitatively, the shifting substellar point due to planetary rotation and/or the stronger zonal winds induced by the rapid rotation would help homogenize temperatures along the equator, irrespective of the surface condition. 
We also note that planets orbiting low-mass stars may attain rotation states other than synchronous rotation (e.g., a 3:2 spin–orbit resonance), in which case the top-of-atmosphere thermal emission becomes more spatially uniform \citep{Turbet+2016,BraamAngerhausen2025}.
The quantitative relationship between temperature gradients and factors such as rotation rate, atmospheric properties, and surface ocean presence remains an open question for future investigation. 

The spectral type of the host star also affects the SNR. 
Due to the larger star-to-planet contrast, thermal emissions of planets around G-type stars would be particularly noisy at wavelengths shorter than the thermal emission peak \citep[e.g.,][Figure 5]{LIFE2}. 
The lack of robust data in the Wien regime of the planetary emission makes it harder to characterize the skewed continuum resulting from the horizontal temperature gradient.

\subsection{Biases in interpreting snapshot spectra of a 3D planet with a 1D model}
\label{ss:dis_Bias1D}

The effects of horizontal temperature gradients on disk-averaged thermal emission spectra, as discussed in Section~\ref{ss:result_snapshot}, may introduce biases in atmospheric retrievals if interpreted using 1D models. 
In particular, the decreasing brightness temperature toward longer wavelengths, caused by surface temperature gradients (Figure~\ref{fig:snapshot_Nc1000mb_1D3D}), can mimic the absorption features of water vapor beyond $>\,10$~$\mu $m. 
The 6 \micron{} H$_2$O feature is more robust against 3D effects. However, spectra in the 6 \micron{} region may not always be well constrained, particularly for planets orbiting solar-type stars, due to contamination from stellar leakage. 
If the nulling efficiency is lower than assumed here, it could substantially affect the robust detection of 6~\micron{} H$_2$O bands even for later-type stars. 
Detection of H$_2$O relying on the absorption band at $> 10$~\micron{} should be interpreted with caution. 

Distortions in the shapes of absorption bands can also introduce retrieval biases.
For example, a narrower CO$_2$ feature caused by three-dimensional atmospheric structure (Figure~\ref{fig:snapshot_Nc1000mb_1D3D}), when observed at low SNR, could result in an underestimation of CO$_2$ abundance due to the reduced equivalent width or an overestimation of the vertical temperature gradient.
In both cases, complementary information from phase variations will assist in discriminating between the scenarios.

\section{Summary}
\label{s:summary}

This paper investigated the effects of 3D temperature structures on the thermal emission spectra of directly imaged, potentially habitable planets.
We adopted Teegarden’s Star b as a benchmark target for future direct imaging observations with \LIFE{}, and simulated atmospheric structures for various scenarios, with and without a global surface ocean, using the ROCKE-3D GCM.
The GCM outputs were then used to simulate geometry-dependent thermal emission spectra as would be observed by \LIFE{}. 

The GCM simulations confirm that planets globally covered by oceans exhibit suppressed day–night contrast in thermal emission. 
In contrast, planets without global ocean develop large temperature gradient even with greenhouse gases and atmospheric water vapor  (up to 1\%). 
These results support the idea that the signatures of horizontal temperature can be considered as an anti-indicator of a global ocean. 

Two types of such signatures have been investigated: phase variation and the spectral shape of the snapshot spectra. 
Phase variations of no-ocean scenarios with $<1$-10~bar atmospheres could be detected with \LIFE{} after 1 days of integration at each of the two orbital phases, assuming four apertures with diameter of 2~m and the nulling baseline up to 100~meter (see Table \ref{tab:lifesim}). 
The clues found in snapshot spectra, monotonically running brightness temperature of the continuum and the peculiar shape of absorption bands in snapshot spectra, would require integration 
longer than the detection of phase variation. 
Conversely, interpreting the spectra of 3D planets using 1D models may introduce substantial biases. 

Similar observations can be applied to other potentially habitable planets in the solar neightborhood, particularly if they orbit closer to the inner edge of habitable zones. 
For planets around M-type stars within 5 parsecs, 1-2 days of integrations would constrain their phase curves. 
The integration time of several days would access a few tens of targets within 10 parsecs, including those which have not been discovered yet. 
Integerations a few times longer would further allow us to investigate the detailed spectral shape originating from the 3D atmospheric structures. 

Although this study focused primarily on the \LIFE{} space interferometer mission, the concept is applicable to other thermal-emission direct imaging efforts. 
The impact of horizontal temperature gradients on thermal emission spectra is also expected to apply to planets orbiting solar-type stars, though interpretation in those cases may be more degenerate. 

This study suggests a pathway for constraining planetary surface conditions from thermal emission measurements aided by GCM simulations. 
Such information complements spectral characterization of atmospheric composition and provides critical context for interpreting it. 
This method is also complementary to other proposed indicators of surface liquid water in scattered-light spectra, such as ocean glint \citep{WilliamsGaidos2008}, ocean color \citep[e.g.,][]{Cowan+2009,Fujii+2010}, or variability in water vapor \citep{Fujii+2013}, which will be accessible with missions like HWO, targeting primarily planets around solar-type (K-type and earlier) stars.

\begin{acknowledgements}
      Numerical computations were carried out on Cray XC50 and XD2000 at the Center for Computational Astrophysics, National Astronomical Observatory of Japan. 
      We are grateful to Kazuyoshi Yamashita and Ryo Kato for assisting with the preparation of the input files for ROCKE-3D. 
      We would like to thank Sascha P. Quanz, Philipp A. Huber, Marrick Braam, and Felix Dannert for their helpful feedback on an earlier draft of this manuscript. 
      We also thank the LIFE Science Team for the fruitful discussions at the meeting. The constructive comments of the anonymous reviewer improved the contents of this manuscript. 
      Y.~F. was supported by JSPS KAKENHI Grant Numbers 18K13601 and 25K01062. 
      Part of this work was supported by the German
      \emph{Deut\-sche For\-schungs\-ge\-mein\-schaft, DFG\/} project
      number Ts~17/2--1. 
\end{acknowledgements}

%
%
\bibliographystyle{aasjournal}
\bibliography{bibtex_yf}

@ARTICLE{Abe+2011,
       author = {{Abe}, Yutaka and {Abe-Ouchi}, Ayako and {Sleep}, Norman H. and {Zahnle}, Kevin J.},
        title = "{Habitable Zone Limits for Dry Planets}",
      journal = {Astrobiology},
         year = 2011,
        month = jun,
       volume = {11},
       number = {5},
        pages = {443-460},
          doi = {10.1089/ast.2010.0545},
       adsurl = {https://ui.adsabs.harvard.edu/abs/2011AsBio..11..443A}
}

@ARTICLE{Allard+2012,
       author = {{Allard}, F. and {Homeier}, D. and {Freytag}, B.},
        title = "{Models of very-low-mass stars, brown dwarfs and exoplanets}",
      journal = {Philosophical Transactions of the Royal Society of London Series A},
         year = 2012,
        month = jun,
       volume = {370},
       number = {1968},
        pages = {2765-2777},
          doi = {10.1098/rsta.2011.0269},
archivePrefix = {arXiv},
       eprint = {1112.3591},
 primaryClass = {astro-ph.SR},
       adsurl = {https://ui.adsabs.harvard.edu/abs/2012RSPTA.370.2765A}
}

@ARTICLE{Amundsen+2017,
       author = {{Amundsen}, David S. and {Tremblin}, Pascal and {Manners}, James and {Baraffe}, Isabelle and {Mayne}, Nathan J.},
        title = "{Treatment of overlapping gaseous absorption with the correlated-k method in hot Jupiter and brown dwarf atmosphere models}",
      journal = {\aap},
         year = 2017,
        month = feb,
       volume = {598},
          eid = {A97},
        pages = {A97},
          doi = {10.1051/0004-6361/201629322},
archivePrefix = {arXiv},
       eprint = {1610.01389},
 primaryClass = {astro-ph.EP},
       adsurl = {https://ui.adsabs.harvard.edu/abs/2017A&A...598A..97A}
}

@ARTICLE{Angerhausen+2024,
       author = {{Angerhausen}, Daniel and {Pidhorodetska}, Daria and {Leung}, Michaela and {Hansen}, Janina and {Alei}, Eleonora and {Dannert}, Felix and {Kammerer}, Jens and {Quanz}, Sascha P. and {Schwieterman}, Edward W. and {The LIFE initiative}},
        title = "{Large Interferometer For Exoplanets (LIFE). XII. The Detectability of Capstone Biosignatures in the Mid-infrared{\textemdash}Sniffing Exoplanetary Laughing Gas and Methylated Halogens}",
      journal = {\aj},
         year = 2024,
        month = mar,
       volume = {167},
       number = {3},
          eid = {128},
        pages = {128},
          doi = {10.3847/1538-3881/ad1f4b},
       adsurl = {https://ui.adsabs.harvard.edu/abs/2024AJ....167..128A}
}

@ARTICLE{Arney+2016,
       author = {{Arney}, Giada and {Domagal-Goldman}, Shawn D. and {Meadows}, Victoria S. and {Wolf}, Eric T. and {Schwieterman}, Edward and {Charnay}, Benjamin and {Claire}, Mark and {H{\'e}brard}, Eric and {Trainer}, Melissa G.},
        title = "{The Pale Orange Dot: The Spectrum and Habitability of Hazy Archean Earth}",
      journal = {Astrobiology},
         year = 2016,
        month = nov,
       volume = {16},
       number = {11},
        pages = {873-899},
          doi = {10.1089/ast.2015.1422},
archivePrefix = {arXiv},
       eprint = {1610.04515},
 primaryClass = {astro-ph.EP},
       adsurl = {https://ui.adsabs.harvard.edu/abs/2016AsBio..16..873A}
}

@ARTICLE{Arney2019,
       author = {{Arney}, Giada N.},
        title = "{The K Dwarf Advantage for Biosignatures on Directly Imaged Exoplanets}",
      journal = {\apjl},
         year = 2019,
        month = mar,
       volume = {873},
       number = {1},
          eid = {L7},
        pages = {L7},
          doi = {10.3847/2041-8213/ab0651},
archivePrefix = {arXiv},
       eprint = {2001.10458},
 primaryClass = {astro-ph.EP},
       adsurl = {https://ui.adsabs.harvard.edu/abs/2019ApJ...873L...7A}
}

@ARTICLE{Barnes+2017,
       author = {{Barnes}, Rory},
        title = "{Tidal locking of habitable exoplanets}",
      journal = {Celestial Mechanics and Dynamical Astronomy},
         year = 2017,
        month = dec,
       volume = {129},
       number = {4},
        pages = {509-536},
          doi = {10.1007/s10569-017-9783-7},
archivePrefix = {arXiv},
       eprint = {1708.02981},
 primaryClass = {astro-ph.EP},
       adsurl = {https://ui.adsabs.harvard.edu/abs/2017CeMDA.129..509B}
}

@ARTICLE{Boukrouche+2024,
       author = {{Boukrouche}, Ryan and {Caballero}, Rodrigo and {Janson}, Markus},
        title = "{The impact of water clouds on the prospective emission spectrum of Teegarden's Star b as observed by LIFE}",
      journal = {arXiv e-prints},
         year = 2024,
        month = nov,
          eid = {arXiv:2411.07922},
        pages = {arXiv:2411.07922},
          doi = {10.48550/arXiv.2411.07922},
archivePrefix = {arXiv},
       eprint = {2411.07922},
 primaryClass = {astro-ph.EP},
       adsurl = {https://ui.adsabs.harvard.edu/abs/2024arXiv241107922B}
}

@ARTICLE{Braam+2025,
       author = {{Braam}, Marrick and {Palmer}, Paul I. and {Decin}, Leen and {Mayne}, Nathan J. and {Manners}, James and {Rugheimer}, Sarah},
        title = "{Earth-like Exoplanets in Spin{\textendash}Orbit Resonances: Climate Dynamics, 3D Atmospheric Chemistry, and Observational Signatures}",
      journal = {\psj},
         year = 2025,
        month = jan,
       volume = {6},
       number = {1},
          eid = {5},
        pages = {5},
          doi = {10.3847/PSJ/ad9565},
archivePrefix = {arXiv},
       eprint = {2410.19108},
 primaryClass = {astro-ph.EP},
       adsurl = {https://ui.adsabs.harvard.edu/abs/2025PSJ.....6....5B}
}

@ARTICLE{BraamAngerhausen2025,
       author = {{Braam}, Marrick and {Angerhausen}, Daniel},
        title = "{Observing spatial and temporal variations in the atmospheric chemistry of rocky exoplanets: prospects for mid-infrared spectroscopy}",
      journal = {arXiv e-prints},
         year = 2025,
        month = dec,
          eid = {arXiv:2512.16619},
        pages = {arXiv:2512.16619},
          doi = {10.48550/arXiv.2512.16619},
archivePrefix = {arXiv},
       eprint = {2512.16619},
 primaryClass = {astro-ph.EP},
       adsurl = {https://ui.adsabs.harvard.edu/abs/2025arXiv251216619B}
}

@ARTICLE{Colose+2021,
       author = {{Colose}, Christopher M. and {Haqq-Misra}, Jacob and {Wolf}, Eric T. and {Del Genio}, Anthony D. and {Barnes}, Rory and {Way}, Michael J. and {Ruedy}, Reto},
        title = "{Effects of Spin-Orbit Resonances and Tidal Heating on the Inner Edge of the Habitable Zone}",
      journal = {\apj},
         year = 2021,
        month = nov,
       volume = {921},
       number = {1},
          eid = {25},
        pages = {25},
          doi = {10.3847/1538-4357/ac135c},
archivePrefix = {arXiv},
       eprint = {2012.07996},
 primaryClass = {astro-ph.EP},
       adsurl = {https://ui.adsabs.harvard.edu/abs/2021ApJ...921...25C}
}

@ARTICLE{Cowan+2009,
       author = {{Cowan}, Nicolas B. and {Agol}, Eric and {Meadows}, Victoria S. and {Robinson}, Tyler and {Livengood}, Timothy A. and {Deming}, Drake and {Lisse}, Carey M. and {A'Hearn}, Michael F. and {Wellnitz}, Dennis D. and {Seager}, Sara and {Charbonneau}, David and {EPOXI Team}},
        title = "{Alien Maps of an Ocean-bearing World}",
      journal = {\apj},
         year = 2009,
        month = aug,
       volume = {700},
       number = {2},
        pages = {915-923},
          doi = {10.1088/0004-637X/700/2/915},
archivePrefix = {arXiv},
       eprint = {0905.3742},
 primaryClass = {astro-ph.EP},
       adsurl = {https://ui.adsabs.harvard.edu/abs/2009ApJ...700..915C}
}

@ARTICLE{DelGenio+2019,
       author = {{Del Genio}, Anthony D. and {Way}, Michael J. and {Amundsen}, David S. and {Aleinov}, Igor and {Kelley}, Maxwell and {Kiang}, Nancy Y. and {Clune}, Thomas L.},
        title = "{Habitable Climate Scenarios for Proxima Centauri b with a Dynamic Ocean}",
      journal = {Astrobiology},
         year = 2019,
        month = jan,
       volume = {19},
       number = {1},
        pages = {99-125},
          doi = {10.1089/ast.2017.1760},
       adsurl = {https://ui.adsabs.harvard.edu/abs/2019AsBio..19...99D}
}

@ARTICLE{DingPierrehumbert2018,
       author = {{Ding}, F. and {Pierrehumbert}, R.~T.},
        title = "{Global or Local Pure Condensible Atmospheres: Importance of Horizontal Latent Heat Transport}",
      journal = {\apj},
         year = 2018,
        month = nov,
       volume = {867},
       number = {1},
          eid = {54},
        pages = {54},
          doi = {10.3847/1538-4357/aae38c},
archivePrefix = {arXiv},
       eprint = {1809.10849},
 primaryClass = {astro-ph.EP},
       adsurl = {https://ui.adsabs.harvard.edu/abs/2018ApJ...867...54D}
}

@ARTICLE{DingPierrehumbert2020,
       author = {{Ding}, Feng and {Pierrehumbert}, Raymond T.},
        title = "{The Phase-curve Signature of Condensible Water-rich Atmospheres on Slowly Rotating Tidally Locked Exoplanets}",
      journal = {\apjl},
         year = 2020,
        month = oct,
       volume = {901},
       number = {2},
          eid = {L33},
        pages = {L33},
          doi = {10.3847/2041-8213/abb941},
archivePrefix = {arXiv},
       eprint = {2009.13638},
 primaryClass = {astro-ph.EP},
       adsurl = {https://ui.adsabs.harvard.edu/abs/2020ApJ...901L..33D}
}

@ARTICLE{Dreizler+2024,
       author = {{Dreizler}, S. and {Luque}, R. and {Ribas}, I. and {Koseleva}, V. and {Ruh}, H.~L. and {Nagel}, E. and {Pozuelos}, F.~J. and {Zechmeister}, M. and {Reiners}, A. and {Caballero}, J.~A. and {Amado}, P.~J. and {B{\'e}jar}, V.~J.~S. and {Bean}, J.~L. and {Brady}, M. and {Cifuentes}, C. and {Gillon}, M. and {Hatzes}, A.~P. and {Henning}, Th. and {Kasper}, D. and {Montes}, D. and {Morales}, J.~C. and {Murray}, C.~A. and {Pall{\'e}}, E. and {Quirrenbach}, A. and {Seifahrt}, A. and {Schweitzer}, A. and {St{\"u}rmer}, J. and {Stef{\'a}nsson}, G. and {Linares}, J.~I. Vico},
        title = "{Teegarden's Star revisited. A nearby planetary system with at least three planets}",
      journal = {\aap},
         year = 2024,
        month = apr,
       volume = {684},
          eid = {A117},
        pages = {A117},
          doi = {10.1051/0004-6361/202348033},
archivePrefix = {arXiv},
       eprint = {2402.00923},
 primaryClass = {astro-ph.EP},
       adsurl = {https://ui.adsabs.harvard.edu/abs/2024A&A...684A.117D}
}

@ARTICLE{DressingCharbonneau2015,
       author = {{Dressing}, Courtney D. and {Charbonneau}, David},
        title = "{The Occurrence of Potentially Habitable Planets Orbiting M Dwarfs Estimated from the Full Kepler Dataset and an Empirical Measurement of the Detection Sensitivity}",
      journal = {\apj},
         year = 2015,
        month = jul,
       volume = {807},
       number = {1},
          eid = {45},
        pages = {45},
          doi = {10.1088/0004-637X/807/1/45},
archivePrefix = {arXiv},
       eprint = {1501.01623},
 primaryClass = {astro-ph.EP},
       adsurl = {https://ui.adsabs.harvard.edu/abs/2015ApJ...807...45D}
}

@ARTICLE{Edwards1996,
       author = {{Edwards}, J.~M.},
        title = "{Efficient Calculation of Infrared Fluxes and Cooling Rates Using the Two-Stream Equations.}",
      journal = {Journal of the Atmospheric Sciences},
         year = 1996,
        month = jul,
       volume = {53},
       number = {13},
        pages = {1921-1932},
          doi = {10.1175/1520-0469(1996)053<1921:ECOIFA>2.0.CO;2},
       adsurl = {https://ui.adsabs.harvard.edu/abs/1996JAtS...53.1921E}
}

@ARTICLE{EdwardsSlingo1996,
       author = {{Edwards}, J.~M. and {Slingo}, A.},
        title = "{Studies with a flexible new radiation code. I: Choosing a configuration for a large-scale model}",
      journal = {Quarterly Journal of the Royal Meteorological Society},
         year = 1996,
        month = apr,
       volume = {122},
       number = {531},
        pages = {689-719},
          doi = {10.1002/qj.49712253107},
       adsurl = {https://ui.adsabs.harvard.edu/abs/1996QJRMS.122..689E}
}

@ARTICLE{Frouin+1996,
       author = {{Frouin}, Robert and {Schwindling}, Myriam and {Deschamps}, Pierre-Yves},
        title = "{Spectral reflectance of sea foam in the visible and near-infrared: In situ measurements and remote sensing implications}",
      journal = {\jgr},
         year = 1996,
        month = jun,
       volume = {101},
       number = {C6},
        pages = {14,361-14,371},
          doi = {10.1029/96JC00629},
       adsurl = {https://ui.adsabs.harvard.edu/abs/1996JGR...10114361F}
}

@ARTICLE{Fujii+2010,
       author = {{Fujii}, Yuka and {Kawahara}, Hajime and {Suto}, Yasushi and {Taruya}, Atsushi and {Fukuda}, Satoru and {Nakajima}, Teruyuki and {Turner}, Edwin L.},
        title = "{Colors of a Second Earth: Estimating the Fractional Areas of Ocean, Land, and Vegetation of Earth-like Exoplanets}",
      journal = {\apj},
         year = 2010,
        month = jun,
       volume = {715},
       number = {2},
        pages = {866-880},
          doi = {10.1088/0004-637X/715/2/866},
archivePrefix = {arXiv},
       eprint = {0911.5621},
 primaryClass = {astro-ph.EP},
       adsurl = {https://ui.adsabs.harvard.edu/abs/2010ApJ...715..866F}
}

@ARTICLE{Fujii+2013,
       author = {{Fujii}, Yuka and {Turner}, Edwin L. and {Suto}, Yasushi},
        title = "{Variability of Water and Oxygen Absorption Bands in the Disk-integrated Spectra of Earth}",
      journal = {\apj},
         year = 2013,
        month = mar,
       volume = {765},
       number = {2},
          eid = {76},
        pages = {76},
          doi = {10.1088/0004-637X/765/2/76},
archivePrefix = {arXiv},
       eprint = {1212.6379},
 primaryClass = {astro-ph.EP},
       adsurl = {https://ui.adsabs.harvard.edu/abs/2013ApJ...765...76F}
}

@ARTICLE{Fujii+2014,
       author = {{Fujii}, Yuka and {Kimura}, Jun and {Dohm}, James and {Ohtake}, Makiko},
        title = "{Geology and Photometric Variation of Solar System Bodies with Minor Atmospheres: Implications for Solid Exoplanets}",
      journal = {Astrobiology},
         year = 2014,
        month = sep,
       volume = {14},
       number = {9},
        pages = {753-768},
          doi = {10.1089/ast.2014.1165},
archivePrefix = {arXiv},
       eprint = {1409.1051},
 primaryClass = {astro-ph.EP},
       adsurl = {https://ui.adsabs.harvard.edu/abs/2014AsBio..14..753F}
}

@ARTICLE{Fujii+2017,
       author = {{Fujii}, Yuka and {Del Genio}, Anthony D. and {Amundsen}, David S.},
        title = "{NIR-driven Moist Upper Atmospheres of Synchronously Rotating Temperate Terrestrial Exoplanets}",
      journal = {\apj},
         year = 2017,
        month = oct,
       volume = {848},
       number = {2},
          eid = {100},
        pages = {100},
          doi = {10.3847/1538-4357/aa8955},
archivePrefix = {arXiv},
       eprint = {1704.05878},
 primaryClass = {astro-ph.EP},
       adsurl = {https://ui.adsabs.harvard.edu/abs/2017ApJ...848..100F}
}

@ARTICLE{Fujii+2018,
       author = {{Fujii}, Yuka and {Angerhausen}, Daniel and {Deitrick}, Russell and {Domagal-Goldman}, Shawn and {Grenfell}, John Lee and {Hori}, Yasunori and {Kane}, Stephen R. and {Pall{\'e}}, Enric and {Rauer}, Heike and {Siegler}, Nicholas and {Stapelfeldt}, Karl and {Stevenson}, Kevin B.},
        title = "{Exoplanet Biosignatures: Observational Prospects}",
      journal = {Astrobiology},
         year = 2018,
        month = jun,
       volume = {18},
       number = {6},
        pages = {739-778},
          doi = {10.1089/ast.2017.1733},
archivePrefix = {arXiv},
       eprint = {1705.07098},
 primaryClass = {astro-ph.EP},
       adsurl = {https://ui.adsabs.harvard.edu/abs/2018AsBio..18..739F}
}

@ARTICLE{FraySchmitt2009,
       author = {{Fray}, N. and {Schmitt}, B.},
        title = "{Sublimation of ices of astrophysical interest: A bibliographic review}",
      journal = {\planss},
         year = 2009,
        month = dec,
       volume = {57},
       number = {14-15},
        pages = {2053-2080},
          doi = {10.1016/j.pss.2009.09.011},
       adsurl = {https://ui.adsabs.harvard.edu/abs/2009P&SS...57.2053F}
}

@ARTICLE{Godolt+2015,
       author = {{Godolt}, M. and {Grenfell}, J.~L. and {Hamann-Reinus}, A. and {Kitzmann}, D. and {Kunze}, M. and {Langematz}, U. and {von Paris}, P. and {Patzer}, A.~B.~C. and {Rauer}, H. and {Stracke}, B.},
        title = "{3D climate modeling of Earth-like extrasolar planets orbiting different types of host stars}",
      journal = {\planss},
         year = 2015,
        month = jun,
       volume = {111},
        pages = {62-76},
          doi = {10.1016/j.pss.2015.03.010},
archivePrefix = {arXiv},
       eprint = {1504.01558},
 primaryClass = {astro-ph.EP},
       adsurl = {https://ui.adsabs.harvard.edu/abs/2015P&SS..111...62G}
}

@ARTICLE{GomezLeal+2012,
       author = {{G{\'o}mez-Leal}, I. and {Pall{\'e}}, E. and {Selsis}, F.},
        title = "{Photometric Variability of the Disk-integrated Thermal Emission of the Earth}",
      journal = {\apj},
         year = 2012,
        month = jun,
       volume = {752},
       number = {1},
          eid = {28},
        pages = {28},
          doi = {10.1088/0004-637X/752/1/28},
archivePrefix = {arXiv},
       eprint = {1205.5010},
 primaryClass = {astro-ph.EP},
       adsurl = {https://ui.adsabs.harvard.edu/abs/2012ApJ...752...28G}
}

@ARTICLE{Goody+1989,
       author = {{Goody}, Richard and {West}, Robert and {Chen}, Luke and {Crisp}, David},
        title = "{The correlated-k method for radiation calculations in nonhomogeneous atmospheres.}",
      journal = {\jqsrt},
         year = 1989,
        month = dec,
       volume = {42},
        pages = {539-550},
          doi = {10.1016/0022-4073(89)90044-7},
       adsurl = {https://ui.adsabs.harvard.edu/abs/1989JQSRT..42..539G}
}

@ARTICLE{Haqq-Misra+2018,
       author = {{Haqq-Misra}, Jacob and {Wolf}, Eric. T. and {Joshi}, Manoj and {Zhang}, Xi and {Kopparapu}, Ravi Kumar},
        title = "{Demarcating Circulation Regimes of Synchronously Rotating Terrestrial Planets within the Habitable Zone}",
      journal = {\apj},
         year = 2018,
        month = jan,
       volume = {852},
       number = {2},
          eid = {67},
        pages = {67},
          doi = {10.3847/1538-4357/aa9f1f},
archivePrefix = {arXiv},
       eprint = {1710.00435},
 primaryClass = {astro-ph.EP},
       adsurl = {https://ui.adsabs.harvard.edu/abs/2018ApJ...852...67H}
}

@ARTICLE{HellerArmstrong2014,
       author = {{Heller}, Ren{\'e} and {Armstrong}, John},
        title = "{Superhabitable Worlds}",
      journal = {Astrobiology},
         year = 2014,
        month = jan,
       volume = {14},
       number = {1},
        pages = {50-66},
          doi = {10.1089/ast.2013.1088},
archivePrefix = {arXiv},
       eprint = {1401.2392},
 primaryClass = {astro-ph.EP},
       adsurl = {https://ui.adsabs.harvard.edu/abs/2014AsBio..14...50H}
}

@ARTICLE{HenryJao2024,
       author = {{Henry}, Todd J. and {Jao}, Wei-Chun},
        title = "{The Character of M Dwarfs}",
      journal = {\araa},
         year = 2024,
        month = sep,
       volume = {62},
       number = {1},
        pages = {593-633},
          doi = {10.1146/annurev-astro-052722-102740},
       adsurl = {https://ui.adsabs.harvard.edu/abs/2024ARA&A..62..593H}
}

@ARTICLE{Hu+2012,
       author = {{Hu}, Renyu and {Ehlmann}, Bethany L. and {Seager}, Sara},
        title = "{Theoretical Spectra of Terrestrial Exoplanet Surfaces}",
      journal = {\apj},
         year = 2012,
        month = jun,
       volume = {752},
       number = {1},
          eid = {7},
        pages = {7},
          doi = {10.1088/0004-637X/752/1/7},
archivePrefix = {arXiv},
       eprint = {1204.1544},
 primaryClass = {astro-ph.EP},
       adsurl = {https://ui.adsabs.harvard.edu/abs/2012ApJ...752....7H}
}

@ARTICLE{Joshi+1997,
       author = {{Joshi}, M.~M. and {Haberle}, R.~M. and {Reynolds}, R.~T.},
        title = "{Simulations of the Atmospheres of Synchronously Rotating Terrestrial Planets Orbiting M Dwarfs: Conditions for Atmospheric Collapse and the Implications for Habitability}",
      journal = {\icarus},
         year = 1997,
        month = oct,
       volume = {129},
       number = {2},
        pages = {450-465},
          doi = {10.1006/icar.1997.5793},
       adsurl = {https://ui.adsabs.harvard.edu/abs/1997Icar..129..450J}
}

@ARTICLE{Kaltenegger+2010,
       author = {{Kaltenegger}, L. and {Henning}, W.~G. and {Sasselov}, D.~D.},
        title = "{Detecting Volcanism on Extrasolar Planets}",
      journal = {\aj},
         year = 2010,
        month = nov,
       volume = {140},
       number = {5},
        pages = {1370-1380},
          doi = {10.1088/0004-6256/140/5/1370},
archivePrefix = {arXiv},
       eprint = {1009.1355},
 primaryClass = {astro-ph.EP},
       adsurl = {https://ui.adsabs.harvard.edu/abs/2010AJ....140.1370K}
}

@ARTICLE{KalteneggerSasselov2010,
       author = {{Kaltenegger}, L. and {Sasselov}, D.},
        title = "{Detecting Planetary Geochemical Cycles on Exoplanets: Atmospheric Signatures and the Case of SO$_{2}$}",
      journal = {\apj},
         year = 2010,
        month = jan,
       volume = {708},
       number = {2},
        pages = {1162-1167},
          doi = {10.1088/0004-637X/708/2/1162},
archivePrefix = {arXiv},
       eprint = {0906.2193},
 primaryClass = {astro-ph.EP},
       adsurl = {https://ui.adsabs.harvard.edu/abs/2010ApJ...708.1162K}
}

@ARTICLE{KammererQuanz2018,
       author = {{Kammerer}, Jens and {Quanz}, Sascha P.},
        title = "{Simulating the exoplanet yield of a space-based mid-infrared interferometer based on Kepler statistics}",
      journal = {\aap},
         year = 2018,
        month = jan,
       volume = {609},
          eid = {A4},
        pages = {A4},
          doi = {10.1051/0004-6361/201731254},
archivePrefix = {arXiv},
       eprint = {1707.06820},
 primaryClass = {astro-ph.EP},
       adsurl = {https://ui.adsabs.harvard.edu/abs/2018A&A...609A...4K}
}

@ARTICLE{KaspiShowman2015,
       author = {{Kaspi}, Yohai and {Showman}, Adam P.},
        title = "{Atmospheric Dynamics of Terrestrial Exoplanets over a Wide Range of Orbital and Atmospheric Parameters}",
      journal = {\apj},
         year = 2015,
        month = may,
       volume = {804},
       number = {1},
          eid = {60},
        pages = {60},
          doi = {10.1088/0004-637X/804/1/60},
archivePrefix = {arXiv},
       eprint = {1407.6349},
 primaryClass = {astro-ph.EP},
       adsurl = {https://ui.adsabs.harvard.edu/abs/2015ApJ...804...60K}
}

@ARTICLE{Knutson+2007,
       author = {{Knutson}, Heather A. and {Charbonneau}, David and {Allen}, Lori E. and {Fortney}, Jonathan J. and {Agol}, Eric and {Cowan}, Nicolas B. and {Showman}, Adam P. and {Cooper}, Curtis S. and {Megeath}, S. Thomas},
        title = "{A map of the day-night contrast of the extrasolar planet HD 189733b}",
      journal = {\nat},
         year = 2007,
        month = may,
       volume = {447},
       number = {7141},
        pages = {183-186},
          doi = {10.1038/nature05782},
archivePrefix = {arXiv},
       eprint = {0705.0993},
 primaryClass = {astro-ph},
       adsurl = {https://ui.adsabs.harvard.edu/abs/2007Natur.447..183K}
}

@ARTICLE{Kodama+2018,
       author = {{Kodama}, T. and {Nitta}, A. and {Genda}, H. and {Takao}, Y. and {O'ishi}, R. and {Abe-Ouchi}, A. and {Abe}, Y.},
        title = "{Dependence of the Onset of the Runaway Greenhouse Effect on the Latitudinal Surface Water Distribution of Earth-Like Planets}",
      journal = {Journal of Geophysical Research (Planets)},
         year = 2018,
        month = feb,
       volume = {123},
       number = {2},
        pages = {559-574},
          doi = {10.1002/2017JE005383},
archivePrefix = {arXiv},
       eprint = {1801.07202},
 primaryClass = {astro-ph.EP},
       adsurl = {https://ui.adsabs.harvard.edu/abs/2018JGRE..123..559K}
}

@ARTICLE{KomacekAbbot2019,
       author = {{Komacek}, Thaddeus D. and {Abbot}, Dorian S.},
        title = "{The Atmospheric Circulation and Climate of Terrestrial Planets Orbiting Sun-like and M Dwarf Stars over a Broad Range of Planetary Parameters}",
      journal = {\apj},
         year = 2019,
        month = feb,
       volume = {871},
       number = {2},
          eid = {245},
        pages = {245},
          doi = {10.3847/1538-4357/aafb33},
archivePrefix = {arXiv},
       eprint = {1901.00567},
 primaryClass = {astro-ph.EP},
       adsurl = {https://ui.adsabs.harvard.edu/abs/2019ApJ...871..245K}
}

@ARTICLE{Konrad+2024,
       author = {{Konrad}, Bj{\"o}rn S. and {Quanz}, Sascha P. and {Alei}, Eleonora and {Wordsworth}, Robin},
        title = "{Pursuing Truth: Improving Retrievals on Mid-infrared Exo-Earth Spectra with Physically Motivated Water Abundance Profiles and Cloud Models}",
      journal = {\apj},
         year = 2024,
        month = nov,
       volume = {975},
       number = {1},
          eid = {13},
        pages = {13},
          doi = {10.3847/1538-4357/ad74f7},
archivePrefix = {arXiv},
       eprint = {2408.09129},
 primaryClass = {astro-ph.EP},
       adsurl = {https://ui.adsabs.harvard.edu/abs/2024ApJ...975...13K}
}

@ARTICLE{KollAbbot2015,
       author = {{Koll}, Daniel D.~B. and {Abbot}, Dorian S.},
        title = "{Deciphering Thermal Phase Curves of Dry, Tidally Locked Terrestrial Planets}",
      journal = {\apj},
         year = 2015,
        month = mar,
       volume = {802},
       number = {1},
          eid = {21},
        pages = {21},
          doi = {10.1088/0004-637X/802/1/21},
archivePrefix = {arXiv},
       eprint = {1412.8216},
 primaryClass = {astro-ph.EP},
       adsurl = {https://ui.adsabs.harvard.edu/abs/2015ApJ...802...21K}
}

@ARTICLE{Kopparapu+2013,
       author = {{Kopparapu}, Ravi Kumar and {Ramirez}, Ramses and {Kasting}, James F. and {Eymet}, Vincent and {Robinson}, Tyler D. and {Mahadevan}, Suvrath and {Terrien}, Ryan C. and {Domagal-Goldman}, Shawn and {Meadows}, Victoria and {Deshpande}, Rohit},
        title = "{Habitable Zones around Main-sequence Stars: New Estimates}",
      journal = {\apj},
         year = 2013,
        month = mar,
       volume = {765},
       number = {2},
          eid = {131},
        pages = {131},
          doi = {10.1088/0004-637X/765/2/131},
archivePrefix = {arXiv},
       eprint = {1301.6674},
 primaryClass = {astro-ph.EP},
       adsurl = {https://ui.adsabs.harvard.edu/abs/2013ApJ...765..131K}
}

@ARTICLE{Kopparapu+2017,
       author = {{Kopparapu}, Ravi kumar and {Wolf}, Eric T. and {Arney}, Giada and {Batalha}, Natasha E. and {Haqq-Misra}, Jacob and {Grimm}, Simon L. and {Heng}, Kevin},
        title = "{Habitable Moist Atmospheres on Terrestrial Planets near the Inner Edge of the Habitable Zone around M Dwarfs}",
      journal = {\apj},
         year = 2017,
        month = aug,
       volume = {845},
       number = {1},
          eid = {5},
        pages = {5},
          doi = {10.3847/1538-4357/aa7cf9},
archivePrefix = {arXiv},
       eprint = {1705.10362},
 primaryClass = {astro-ph.EP},
       adsurl = {https://ui.adsabs.harvard.edu/abs/2017ApJ...845....5K}
}

@ARTICLE{KreidbergLoeb2016,
       author = {{Kreidberg}, Laura and {Loeb}, Abraham},
        title = "{Prospects for Characterizing the Atmosphere of Proxima Centauri b}",
      journal = {\apjl},
         year = 2016,
        month = nov,
       volume = {832},
       number = {1},
          eid = {L12},
        pages = {L12},
          doi = {10.3847/2041-8205/832/1/L12},
archivePrefix = {arXiv},
       eprint = {1608.07345},
 primaryClass = {astro-ph.EP},
       adsurl = {https://ui.adsabs.harvard.edu/abs/2016ApJ...832L..12K}
}

@ARTICLE{Kreidberg+2019,
       author = {{Kreidberg}, Laura and {Koll}, Daniel D.~B. and {Morley}, Caroline and {Hu}, Renyu and {Schaefer}, Laura and {Deming}, Drake and {Stevenson}, Kevin B. and {Dittmann}, Jason and {Vanderburg}, Andrew and {Berardo}, David and {Guo}, Xueying and {Stassun}, Keivan and {Crossfield}, Ian and {Charbonneau}, David and {Latham}, David W. and {Loeb}, Abraham and {Ricker}, George and {Seager}, Sara and {Vanderspek}, Roland},
        title = "{Absence of a thick atmosphere on the terrestrial exoplanet LHS 3844b}",
      journal = {\nat},
         year = 2019,
        month = aug,
       volume = {573},
       number = {7772},
        pages = {87-90},
          doi = {10.1038/s41586-019-1497-4},
archivePrefix = {arXiv},
       eprint = {1908.06834},
 primaryClass = {astro-ph.EP},
       adsurl = {https://ui.adsabs.harvard.edu/abs/2019Natur.573...87K}
}

@ARTICLE{LacisOinas1991,
       author = {{Lacis}, A.~A. and {Oinas}, V.},
        title = "{A description of the correlated-k distribution method for modelling nongray gaseous absorption, thermal emission, and multiple scattering in vertically inhomogeneous atmospheres}",
      journal = {\jgr},
         year = 1991,
        month = may,
       volume = {96},
        pages = {9027-9064},
          doi = {10.1029/90JD01945},
       adsurl = {https://ui.adsabs.harvard.edu/abs/1991JGR....96.9027L}
}

@ARTICLE{Lobo+2023,
       author = {{Lobo}, Ana H. and {Shields}, Aomawa L. and {Palubski}, Igor Z. and {Wolf}, Eric},
        title = "{Terminator Habitability: The Case for Limited Water Availability on M-dwarf Planets}",
      journal = {\apj},
         year = 2023,
        month = mar,
       volume = {945},
       number = {2},
          eid = {161},
        pages = {161},
          doi = {10.3847/1538-4357/aca970},
archivePrefix = {arXiv},
       eprint = {2212.06185},
 primaryClass = {astro-ph.EP},
       adsurl = {https://ui.adsabs.harvard.edu/abs/2023ApJ...945..161L}
}

@ARTICLE{LoboShields2024,
       author = {{Lobo}, Ana H. and {Shields}, Aomawa L.},
        title = "{Climate Regimes across the Habitable Zone: A Comparison of Synchronous Rocky M and K Dwarf Planets}",
      journal = {\apj},
         year = 2024,
        month = sep,
       volume = {972},
       number = {1},
        pages = {71},
          doi = {10.3847/1538-4357/ad58bb},
archivePrefix = {arXiv},
       eprint = {2408.05951},
 primaryClass = {astro-ph.EP},
       adsurl = {https://ui.adsabs.harvard.edu/abs/2024ApJ...972...71L}
}

@ARTICLE{Lellouch+2000,
       author = {{Lellouch}, E. and {Encrenaz}, T. and {de Graauw}, T. and {Erard}, S. and {Morris}, P. and {Crovisier}, J. and {Feuchtgruber}, H. and {Girard}, T. and {Burgdorf}, M.},
        title = "{The 2.4- 45 {\ensuremath{\mu}}m spectrum of Mars observed with the infrared space observatory}",
      journal = {\planss},
         year = 2000,
        month = oct,
       volume = {48},
       number = {12-14},
        pages = {1393-1405},
          doi = {10.1016/S0032-0633(00)00118-5},
       adsurl = {https://ui.adsabs.harvard.edu/abs/2000P&SS...48.1393L}
}

@ARTICLE{Malik+2017,
       author = {{Malik}, Matej and {Grosheintz}, Luc and {Mendon{\c{c}}a}, Jo{\~a}o M. and {Grimm}, Simon L. and {Lavie}, Baptiste and {Kitzmann}, Daniel and {Tsai}, Shang-Min and {Burrows}, Adam and {Kreidberg}, Laura and {Bedell}, Megan and {Bean}, Jacob L. and {Stevenson}, Kevin B. and {Heng}, Kevin},
        title = "{HELIOS: An Open-source, GPU-accelerated Radiative Transfer Code for Self-consistent Exoplanetary Atmospheres}",
      journal = {\aj},
         year = 2017,
        month = feb,
       volume = {153},
       number = {2},
          eid = {56},
        pages = {56},
          doi = {10.3847/1538-3881/153/2/56},
archivePrefix = {arXiv},
       eprint = {1606.05474},
 primaryClass = {astro-ph.EP},
       adsurl = {https://ui.adsabs.harvard.edu/abs/2017AJ....153...56M}
}

@ARTICLE{Malik+2019a,
       author = {{Malik}, Matej and {Kitzmann}, Daniel and {Mendon{\c{c}}a}, Jo{\~a}o M. and {Grimm}, Simon L. and {Marleau}, Gabriel-Dominique and {Linder}, Esther F. and {Tsai}, Shang-Min and {Heng}, Kevin},
        title = "{Self-luminous and Irradiated Exoplanetary Atmospheres Explored with HELIOS}",
      journal = {\aj},
         year = 2019,
        month = may,
       volume = {157},
       number = {5},
          eid = {170},
        pages = {170},
          doi = {10.3847/1538-3881/ab1084},
archivePrefix = {arXiv},
       eprint = {1903.06794},
 primaryClass = {astro-ph.EP},
       adsurl = {https://ui.adsabs.harvard.edu/abs/2019AJ....157..170M}
}

@ARTICLE{Malik+2019b,
       author = {{Malik}, Matej and {Kempton}, Eliza M. -R. and {Koll}, Daniel D.~B. and {Mansfield}, Megan and {Bean}, Jacob L. and {Kite}, Edwin},
        title = "{Analyzing Atmospheric Temperature Profiles and Spectra of M Dwarf Rocky Planets}",
      journal = {\apj},
         year = 2019,
        month = dec,
       volume = {886},
       number = {2},
          eid = {142},
        pages = {142},
          doi = {10.3847/1538-4357/ab4a05},
archivePrefix = {arXiv},
       eprint = {1907.13135},
 primaryClass = {astro-ph.EP},
       adsurl = {https://ui.adsabs.harvard.edu/abs/2019ApJ...886..142M}
}

@ARTICLE{Mettler+2020,
       author = {{Mettler}, Jean-No{\"e}l and {Quanz}, Sascha P. and {Helled}, Ravit},
        title = "{Earth as an Exoplanet. I. Time Variable Thermal Emission Using Spatially Resolved Moderate Imaging Spectroradiometer Data}",
      journal = {\aj},
         year = 2020,
        month = dec,
       volume = {160},
       number = {6},
          eid = {246},
        pages = {246},
          doi = {10.3847/1538-3881/abbc15},
archivePrefix = {arXiv},
       eprint = {2010.02589},
 primaryClass = {astro-ph.EP},
       adsurl = {https://ui.adsabs.harvard.edu/abs/2020AJ....160..246M}
}

@ARTICLE{Mettler+2023,
       author = {{Mettler}, Jean-No{\"e}l and {Quanz}, Sascha P. and {Helled}, Ravit and {Olson}, Stephanie L. and {Schwieterman}, Edward W.},
        title = "{Earth as an Exoplanet. II. Earth's Time-variable Thermal Emission and Its Atmospheric Seasonality of Bioindicators}",
      journal = {\apj},
         year = 2023,
        month = apr,
       volume = {946},
       number = {2},
          eid = {82},
        pages = {82},
          doi = {10.3847/1538-4357/acbe3c},
archivePrefix = {arXiv},
       eprint = {2210.05414},
 primaryClass = {astro-ph.EP},
       adsurl = {https://ui.adsabs.harvard.edu/abs/2023ApJ...946...82M}
}

@INPROCEEDINGS{Muller+2020,
       author = {{M{\"u}ller}, Thomas G. and {Burgdorf}, Martin J. and {Buehler}, Stefan A. and {Prange}, Marc},
        title = "{Thermophysical model of the Moon from 3.7 to 15 {\ensuremath{\mu}}m}",
    booktitle = {European Planetary Science Congress},
         year = 2020,
        month = sep,
          eid = {EPSC2020-586},
        pages = {EPSC2020-586},
          doi = {10.5194/epsc2020-586},
       adsurl = {https://ui.adsabs.harvard.edu/abs/2020EPSC...14..586M}
}

@INPROCEEDINGS{Quanz+2018,
       author = {{Quanz}, Sascha P. and {Kammerer}, Jens and {Defr{\`e}re}, Denis and {Absil}, Olivier and {Glauser}, Adrian M. and {Kitzmann}, Daniel},
        title = "{Exoplanet science with a space-based mid-infrared nulling interferometer}",
    booktitle = {Optical and Infrared Interferometry and Imaging VI},
         year = 2018,
       editor = {{Creech-Eakman}, Michelle J. and {Tuthill}, Peter G. and {M{\'e}rand}, Antoine},
       series = {Society of Photo-Optical Instrumentation Engineers (SPIE) Conference Series},
       volume = {10701},
        month = jul,
          eid = {107011I},
        pages = {107011I},
          doi = {10.1117/12.2312051},
archivePrefix = {arXiv},
       eprint = {1807.06088},
 primaryClass = {astro-ph.IM},
       adsurl = {https://ui.adsabs.harvard.edu/abs/2018SPIE10701E..1IQ}
}

@ARTICLE{Robinson+2011,
       author = {{Robinson}, Tyler D. and {Meadows}, Victoria S. and {Crisp}, David and {Deming}, Drake and {A'Hearn}, Michael F. and {Charbonneau}, David and {Livengood}, Timothy A. and {Seager}, Sara and {Barry}, Richard K. and {Hearty}, Thomas and {Hewagama}, Tilak and {Lisse}, Carey M. and {McFadden}, Lucy A. and {Wellnitz}, Dennis D.},
        title = "{Earth as an Extrasolar Planet: Earth Model Validation Using EPOXI Earth Observations}",
      journal = {Astrobiology},
         year = 2011,
        month = jun,
       volume = {11},
       number = {5},
        pages = {393-408},
          doi = {10.1089/ast.2011.0642},
       adsurl = {https://ui.adsabs.harvard.edu/abs/2011AsBio..11..393R}
}

@ARTICLE{Robinson2011,
       author = {{Robinson}, Tyler D.},
        title = "{Modeling the Infrared Spectrum of the Earth-Moon System: Implications for the Detection and Characterization of Earthlike Extrasolar Planets and Their Moonlike Companions}",
      journal = {\apj},
         year = 2011,
        month = nov,
       volume = {741},
       number = {1},
          eid = {51},
        pages = {51},
          doi = {10.1088/0004-637X/741/1/51},
archivePrefix = {arXiv},
       eprint = {1110.3744},
 primaryClass = {astro-ph.EP},
       adsurl = {https://ui.adsabs.harvard.edu/abs/2011ApJ...741...51R}
}

@INCOLLECTION{Robinson2018,
       author = {{Robinson}, Tyler D.},
        title = "{Characterizing Exoplanet Habitability}",
    booktitle = {Handbook of Exoplanets},
         year = 2018,
       editor = {{Deeg}, Hans J. and {Belmonte}, Juan Antonio},
          eid = {67},
        pages = {67},
          doi = {10.1007/978-3-319-55333-7_67},
       adsurl = {https://ui.adsabs.harvard.edu/abs/2018haex.bookE..67R}
}

@ARTICLE{Rothman+2013,
       author = {{Rothman}, L.~S. and {Gordon}, I.~E. and {Babikov}, Y. and {Barbe}, A. and {Chris Benner}, D. and {Bernath}, P.~F. and {Birk}, M. and {Bizzocchi}, L. and {Boudon}, V. and {Brown}, L.~R. and {Campargue}, A. and {Chance}, K. and {Cohen}, E.~A. and {Coudert}, L.~H. and {Devi}, V.~M. and {Drouin}, B.~J. and {Fayt}, A. and {Flaud}, J. -M. and {Gamache}, R.~R. and {Harrison}, J.~J. and {Hartmann}, J. -M. and {Hill}, C. and {Hodges}, J.~T. and {Jacquemart}, D. and {Jolly}, A. and {Lamouroux}, J. and {Le Roy}, R.~J. and {Li}, G. and {Long}, D.~A. and {Lyulin}, O.~M. and {Mackie}, C.~J. and {Massie}, S.~T. and {Mikhailenko}, S. and {M{\"u}ller}, H.~S.~P. and {Naumenko}, O.~V. and {Nikitin}, A.~V. and {Orphal}, J. and {Perevalov}, V. and {Perrin}, A. and {Polovtseva}, E.~R. and {Richard}, C. and {Smith}, M.~A.~H. and {Starikova}, E. and {Sung}, K. and {Tashkun}, S. and {Tennyson}, J. and {Toon}, G.~C. and {Tyuterev}, Vl. G. and {Wagner}, G.},
        title = "{The HITRAN2012 molecular spectroscopic database}",
      journal = {\jqsrt},
         year = 2013,
        month = nov,
       volume = {130},
        pages = {4-50},
          doi = {10.1016/j.jqsrt.2013.07.002},
       adsurl = {https://ui.adsabs.harvard.edu/abs/2013JQSRT.130....4R}
}

@ARTICLE{RugheimerKaltenegger2018,
       author = {{Rugheimer}, S. and {Kaltenegger}, L.},
        title = "{Spectra of Earth-like Planets through Geological Evolution around FGKM Stars}",
      journal = {\apj},
         year = 2018,
        month = feb,
       volume = {854},
       number = {1},
          eid = {19},
        pages = {19},
          doi = {10.3847/1538-4357/aaa47a},
archivePrefix = {arXiv},
       eprint = {1712.10027},
 primaryClass = {astro-ph.EP},
       adsurl = {https://ui.adsabs.harvard.edu/abs/2018ApJ...854...19R}
}

@ARTICLE{Rugheimer+2015,
       author = {{Rugheimer}, S. and {Kaltenegger}, L. and {Segura}, A. and {Linsky}, J. and {Mohanty}, S.},
        title = "{Effect of UV Radiation on the Spectral Fingerprints of Earth-like Planets Orbiting M Stars}",
      journal = {\apj},
         year = 2015,
        month = aug,
       volume = {809},
       number = {1},
          eid = {57},
        pages = {57},
          doi = {10.1088/0004-637X/809/1/57},
archivePrefix = {arXiv},
       eprint = {1506.07202},
 primaryClass = {astro-ph.EP},
       adsurl = {https://ui.adsabs.harvard.edu/abs/2015ApJ...809...57R}
}

@ARTICLE{Rugheimer+2013,
       author = {{Rugheimer}, Sarah and {Kaltenegger}, Lisa and {Zsom}, Andras and {Segura}, Ant{\'\i}gona and {Sasselov}, Dimitar},
        title = "{Spectral Fingerprints of Earth-like Planets Around FGK Stars}",
      journal = {Astrobiology},
         year = 2013,
        month = mar,
       volume = {13},
       number = {3},
        pages = {251-269},
          doi = {10.1089/ast.2012.0888},
archivePrefix = {arXiv},
       eprint = {1212.2638},
 primaryClass = {astro-ph.EP},
       adsurl = {https://ui.adsabs.harvard.edu/abs/2013AsBio..13..251R}
}

@ARTICLE{Schlawin+2024,
       author = {{Schlawin}, Everett and {Mukherjee}, Sagnick and {Ohno}, Kazumasa and {Bell}, Taylor J. and {Beatty}, Thomas G. and {Greene}, Thomas P. and {Line}, Michael and {Challener}, Ryan C. and {Parmentier}, Vivien and {Fortney}, Jonathan J. and {Rauscher}, Emily and {Wiser}, Lindsey and {Welbanks}, Luis and {Murphy}, Matthew and {Edelman}, Isaac and {Batalha}, Natasha and {Moran}, Sarah E. and {Mehta}, Nishil and {Rieke}, Marcia},
        title = "{Multiple Clues for Dayside Aerosols and Temperature Gradients in WASP-69 b from a Panchromatic JWST Emission Spectrum}",
      journal = {\aj},
         year = 2024,
        month = sep,
       volume = {168},
       number = {3},
          eid = {104},
        pages = {104},
          doi = {10.3847/1538-3881/ad58e0},
archivePrefix = {arXiv},
       eprint = {2406.15543},
 primaryClass = {astro-ph.EP},
       adsurl = {https://ui.adsabs.harvard.edu/abs/2024AJ....168..104S}
}

@ARTICLE{Selsis+2011,
       author = {{Selsis}, F. and {Wordsworth}, R.~D. and {Forget}, F.},
        title = "{Thermal phase curves of nontransiting terrestrial exoplanets. I. Characterizing atmospheres}",
      journal = {\aap},
         year = 2011,
        month = aug,
       volume = {532},
          eid = {A1},
        pages = {A1},
          doi = {10.1051/0004-6361/201116654},
archivePrefix = {arXiv},
       eprint = {1104.4763},
 primaryClass = {astro-ph.EP},
       adsurl = {https://ui.adsabs.harvard.edu/abs/2011A&A...532A...1S}
}

@ARTICLE{Taysum+2024,
       author = {{Taysum}, B. and {van Zelst}, I. and {Grenfell}, J.~L. and {Schreier}, F. and {Cabrera}, J. and {Rauer}, H.},
        title = "{Detectability of biosignatures in warm, water-rich atmospheres}",
      journal = {\aap},
         year = 2024,
        month = dec,
       volume = {692},
          eid = {A82},
        pages = {A82},
          doi = {10.1051/0004-6361/202450549},
archivePrefix = {arXiv},
       eprint = {2412.01266},
 primaryClass = {astro-ph.EP},
       adsurl = {https://ui.adsabs.harvard.edu/abs/2024A&A...692A..82T}
}

@ARTICLE{Tinetti+2005,
       author = {{Tinetti}, Giovanna and {Meadows}, Victoria S. and {Crisp}, David and {Fong },William and {Velusamy}, Thangasamy and {Snively}, Heather},
        title = "{Disk-Averaged Synthetic Spectra of Mars}",
      journal = {Astrobiology},
         year = 2005,
        month = aug,
       volume = {5},
       number = {4},
        pages = {461-482},
          doi = {10.1089/ast.2005.5.461},
archivePrefix = {arXiv},
       eprint = {astro-ph/0408372},
 primaryClass = {astro-ph},
       adsurl = {https://ui.adsabs.harvard.edu/abs/2005AsBio...5..461T}
}

@ARTICLE{Tinetti+2006a,
       author = {{Tinetti}, Giovanna and {Meadows}, Victoria S. and {Crisp}, David and {Fong}, William and {Fishbein}, Evan and {Turnbull}, Margaret and {Bibring}, Jean-Pierre},
        title = "{Detectability of Planetary Characteristics in Disk-Averaged Spectra. I: The Earth Model}",
      journal = {Astrobiology},
         year = 2006,
        month = mar,
       volume = {6},
       number = {1},
        pages = {34-47},
          doi = {10.1089/ast.2006.6.34},
       adsurl = {https://ui.adsabs.harvard.edu/abs/2006AsBio...6...34T}
}

@ARTICLE{Tinetti+2006b,
       author = {{Tinetti}, Giovanna and {Meadows}, Victoria S. and {Crisp}, David and {Kiang}, Nancy Y. and {Kahn}, Brian H. and {Fishbein}, Evan and {Velusamy}, Thangasamy and {Turnbull}, Margaret},
        title = "{Detectability of Planetary Characteristics in Disk-Averaged Spectra II: Synthetic Spectra and Light-Curves of Earth}",
      journal = {Astrobiology},
         year = 2006,
        month = dec,
       volume = {6},
       number = {6},
        pages = {881-900},
          doi = {10.1089/ast.2006.6.881},
       adsurl = {https://ui.adsabs.harvard.edu/abs/2006AsBio...6..881T}
}

@ARTICLE{Turbet+2016,
       author = {{Turbet}, Martin and {Leconte}, J{\'e}r{\'e}my and {Selsis}, Franck and {Bolmont}, Emeline and {Forget}, Fran{\c{c}}ois and {Ribas}, Ignasi and {Raymond}, Sean N. and {Anglada-Escud{\'e}}, Guillem},
        title = "{The habitability of Proxima Centauri b. II. Possible climates and observability}",
      journal = {\aap},
         year = 2016,
        month = dec,
       volume = {596},
          eid = {A112},
        pages = {A112},
          doi = {10.1051/0004-6361/201629577},
archivePrefix = {arXiv},
       eprint = {1608.06827},
 primaryClass = {astro-ph.EP},
       adsurl = {https://ui.adsabs.harvard.edu/abs/2016A&A...596A.112T}
}

@ARTICLE{Turbet+2018,
       author = {{Turbet}, Martin and {Bolmont}, Emeline and {Leconte}, Jeremy and {Forget}, Fran{\c{c}}ois and {Selsis}, Franck and {Tobie}, Gabriel and {Caldas}, Anthony and {Naar}, Joseph and {Gillon}, Micha{\"e}l},
        title = "{Modeling climate diversity, tidal dynamics and the fate of volatiles on TRAPPIST-1 planets}",
      journal = {\aap},
         year = 2018,
        month = may,
       volume = {612},
          eid = {A86},
        pages = {A86},
          doi = {10.1051/0004-6361/201731620},
archivePrefix = {arXiv},
       eprint = {1707.06927},
 primaryClass = {astro-ph.EP},
       adsurl = {https://ui.adsabs.harvard.edu/abs/2018A&A...612A..86T}
}

@ARTICLE{WangYang2022,
       author = {{Wang}, Shuang and {Yang}, Jun},
        title = "{Atmospheric Overturning Circulation on Dry, Tidally Locked Rocky Planets Is Mainly Driven by Radiative Cooling}",
      journal = {\psj},
         year = 2022,
        month = jul,
       volume = {3},
       number = {7},
          eid = {171},
        pages = {171},
          doi = {10.3847/PSJ/ac6d65},
archivePrefix = {arXiv},
       eprint = {2206.07249},
 primaryClass = {astro-ph.EP},
       adsurl = {https://ui.adsabs.harvard.edu/abs/2022PSJ.....3..171W}
}

@ARTICLE{Way+2016,
       author = {{Way}, M.~J. and {Del Genio}, Anthony D. and {Kiang}, Nancy Y. and {Sohl}, Linda E. and {Grinspoon}, David H. and {Aleinov}, Igor and {Kelley}, Maxwell and {Clune}, Thomas},
        title = "{Was Venus the first habitable world of our solar system?}",
      journal = {\grl},
         year = 2016,
        month = aug,
       volume = {43},
       number = {16},
        pages = {8376-8383},
          doi = {10.1002/2016GL069790},
archivePrefix = {arXiv},
       eprint = {1608.00706},
 primaryClass = {astro-ph.EP},
       adsurl = {https://ui.adsabs.harvard.edu/abs/2016GeoRL..43.8376W}
}

@ARTICLE{Way+2017,
       author = {{Way}, M.~J. and {Aleinov}, I. and {Amundsen}, David S. and {Chandler}, M.~A. and {Clune}, T.~L. and {Del Genio}, A.~D. and {Fujii}, Y. and {Kelley}, M. and {Kiang}, N.~Y. and {Sohl}, L. and {Tsigaridis}, K.},
        title = "{Resolving Orbital and Climate Keys of Earth and Extraterrestrial Environments with Dynamics (ROCKE-3D) 1.0: A General Circulation Model for Simulating the Climates of Rocky Planets}",
      journal = {\apjs},
         year = 2017,
        month = jul,
       volume = {231},
       number = {1},
          eid = {12},
        pages = {12},
          doi = {10.3847/1538-4365/aa7a06},
archivePrefix = {arXiv},
       eprint = {1701.02360},
 primaryClass = {astro-ph.EP},
       adsurl = {https://ui.adsabs.harvard.edu/abs/2017ApJS..231...12W}
}

@ARTICLE{Whittaker+2022,
       author = {{Whittaker}, Emily A. and {Malik}, Matej and {Ih}, Jegug and {Kempton}, Eliza M. -R. and {Mansfield}, Megan and {Bean}, Jacob L. and {Kite}, Edwin S. and {Koll}, Daniel D.~B. and {Cronin}, Timothy W. and {Hu}, Renyu},
        title = "{The Detectability of Rocky Planet Surface and Atmosphere Composition with the JWST: The Case of LHS 3844b}",
      journal = {\aj},
         year = 2022,
        month = dec,
       volume = {164},
       number = {6},
          eid = {258},
        pages = {258},
          doi = {10.3847/1538-3881/ac9ab3},
archivePrefix = {arXiv},
       eprint = {2207.08889},
 primaryClass = {astro-ph.EP},
       adsurl = {https://ui.adsabs.harvard.edu/abs/2022AJ....164..258W}
}

@ARTICLE{WilliamsGaidos2008,
       author = {{Williams}, Darren M. and {Gaidos}, Eric},
        title = "{Detecting the glint of starlight on the oceans of distant planets}",
      journal = {\icarus},
         year = 2008,
        month = jun,
       volume = {195},
       number = {2},
        pages = {927-937},
          doi = {10.1016/j.icarus.2008.01.002},
archivePrefix = {arXiv},
       eprint = {0801.1852},
 primaryClass = {astro-ph},
       adsurl = {https://ui.adsabs.harvard.edu/abs/2008Icar..195..927W}
}

@ARTICLE{WiscombeEvans1977,
       author = {{Wiscombe}, W.~J. and {Evans}, J.~W.},
        title = "{Exponential-sum fitting of radiative transmission functions.}",
      journal = {Journal of Computational Physics},
         year = 1977,
        month = aug,
       volume = {24},
        pages = {416-444},
          doi = {10.1016/0021-9991(77)90031-6},
       adsurl = {https://ui.adsabs.harvard.edu/abs/1977JCoPh..24..416W}
}

@ARTICLE{Wolf+2019,
       author = {{Wolf}, E.~T. and {Kopparapu}, R.~K. and {Haqq-Misra}, J.},
        title = "{Simulated Phase-dependent Spectra of Terrestrial Aquaplanets in M Dwarf Systems}",
      journal = {\apj},
         year = 2019,
        month = may,
       volume = {877},
       number = {1},
          eid = {35},
        pages = {35},
          doi = {10.3847/1538-4357/ab184a},
archivePrefix = {arXiv},
       eprint = {1906.02697},
 primaryClass = {astro-ph.EP},
       adsurl = {https://ui.adsabs.harvard.edu/abs/2019ApJ...877...35W}
}

@ARTICLE{Yang+2013,
       author = {{Yang}, Jun and {Cowan}, Nicolas B. and {Abbot}, Dorian S.},
        title = "{Stabilizing Cloud Feedback Dramatically Expands the Habitable Zone of Tidally Locked Planets}",
      journal = {\apjl},
         year = 2013,
        month = jul,
       volume = {771},
       number = {2},
          eid = {L45},
        pages = {L45},
          doi = {10.1088/2041-8205/771/2/L45},
archivePrefix = {arXiv},
       eprint = {1307.0515},
 primaryClass = {astro-ph.EP},
       adsurl = {https://ui.adsabs.harvard.edu/abs/2013ApJ...771L..45Y}
}

@ARTICLE{Yang+2014,
       author = {{Yang}, Jun and {Bou{\'e}}, Gwena{\"e}l and {Fabrycky}, Daniel C. and {Abbot}, Dorian S.},
        title = "{Strong Dependence of the Inner Edge of the Habitable Zone on Planetary Rotation Rate}",
      journal = {\apjl},
         year = 2014,
        month = may,
       volume = {787},
       number = {1},
          eid = {L2},
        pages = {L2},
          doi = {10.1088/2041-8205/787/1/L2},
archivePrefix = {arXiv},
       eprint = {1404.4992},
 primaryClass = {astro-ph.EP},
       adsurl = {https://ui.adsabs.harvard.edu/abs/2014ApJ...787L...2Y}
}

@ARTICLE{Yang+2019,
       author = {{Yang}, Jun and {Abbot}, Dorian S. and {Koll}, Daniel D.~B. and {Hu}, Yongyun and {Showman}, Adam P.},
        title = "{Ocean Dynamics and the Inner Edge of the Habitable Zone for Tidally Locked Terrestrial Planets}",
      journal = {\apj},
         year = 2019,
        month = jan,
       volume = {871},
       number = {1},
          eid = {29},
        pages = {29},
          doi = {10.3847/1538-4357/aaf1a8},
archivePrefix = {arXiv},
       eprint = {1902.02103},
 primaryClass = {astro-ph.EP},
       adsurl = {https://ui.adsabs.harvard.edu/abs/2019ApJ...871...29Y}
}

@ARTICLE{Yang+2023,
       author = {{Yang}, Jun and {Zhang}, Yixiao and {Fu}, Zuntao and {Yan}, Mingyu and {Song}, Xinyi and {Wei}, Mengyu and {Liu}, Jiachen and {Ding}, Feng and {Tan}, Zhihong},
        title = "{Cloud behaviour on tidally locked rocky planets from global high-resolution modelling}",
      journal = {Nature Astronomy},
         year = 2023,
        month = sep,
       volume = {7},
        pages = {1070-1080},
          doi = {10.1038/s41550-023-02015-8},
archivePrefix = {arXiv},
       eprint = {2306.12186},
 primaryClass = {astro-ph.EP},
       adsurl = {https://ui.adsabs.harvard.edu/abs/2023NatAs...7.1070Y}
}

@ARTICLE{Zechmeister+2019,
       author = {{Zechmeister}, M. and {Dreizler}, S. and {Ribas}, I. and {Reiners}, A. and {Caballero}, J.~A. and {Bauer}, F.~F. and {B{\'e}jar}, V.~J.~S. and {Gonz{\'a}lez-Cuesta}, L. and {Herrero}, E. and {Lalitha}, S. and {L{\'o}pez-Gonz{\'a}lez}, M.~J. and {Luque}, R. and {Morales}, J.~C. and {Pall{\'e}}, E. and {Rodr{\'\i}guez}, E. and {Rodr{\'\i}guez L{\'o}pez}, C. and {Tal-Or}, L. and {Anglada-Escud{\'e}}, G. and {Quirrenbach}, A. and {Amado}, P.~J. and {Abril}, M. and {Aceituno}, F.~J. and {Aceituno}, J. and {Alonso-Floriano}, F.~J. and {Ammler-von Eiff}, M. and {Antona Jim{\'e}nez}, R. and {Anwand-Heerwart}, H. and {Arroyo-Torres}, B. and {Azzaro}, M. and {Baroch}, D. and {Barrado}, D. and {Becerril}, S. and {Ben{\'\i}tez}, D. and {Berdi{\~n}as}, Z.~M. and {Bergond}, G. and {Bluhm}, P. and {Brinkm{\"o}ller}, M. and {del Burgo}, C. and {Calvo Ortega}, R. and {Cano}, J. and {Cardona Guill{\'e}n}, C. and {Carro}, J. and {C{\'a}rdenas V{\'a}zquez}, M.~C. and {Casal}, E. and {Casasayas-Barris}, N. and {Casanova}, V. and {Chaturvedi}, P. and {Cifuentes}, C. and {Claret}, A. and {Colom{\'e}}, J. and {Cort{\'e}s-Contreras}, M. and {Czesla}, S. and {D{\'\i}ez-Alonso}, E. and {Dorda}, R. and {Fern{\'a}ndez}, M. and {Fern{\'a}ndez-Mart{\'\i}n}, A. and {Fuhrmeister}, B. and {Fukui}, A. and {Galad{\'\i}-Enr{\'\i}quez}, D. and {Gallardo Cava}, I. and {Garcia de la Fuente}, J. and {Garcia-Piquer}, A. and {Garc{\'\i}a Vargas}, M.~L. and {Gesa}, L. and {G{\'o}ngora Rueda}, J. and {Gonz{\'a}lez-{\'A}lvarez}, E. and {Gonz{\'a}lez Hern{\'a}ndez}, J.~I. and {Gonz{\'a}lez-Peinado}, R. and {Gr{\"o}zinger}, U. and {Gu{\`a}rdia}, J. and {Guijarro}, A. and {de Guindos}, E. and {Hatzes}, A.~P. and {Hauschildt}, P.~H. and {Hedrosa}, R.~P. and {Helmling}, J. and {Henning}, T. and {Hermelo}, I. and {Hern{\'a}ndez Arabi}, R. and {Hern{\'a}ndez Casta{\~n}o}, L. and {Hern{\'a}ndez Otero}, F. and {Hintz}, D. and {Huke}, P. and {Huber}, A. and {Jeffers}, S.~V. and {Johnson}, E.~N. and {de Juan}, E. and {Kaminski}, A. and {Kemmer}, J. and {Kim}, M. and {Klahr}, H. and {Klein}, R. and {Kl{\"u}ter}, J. and {Klutsch}, A. and {Kossakowski}, D. and {K{\"u}rster}, M. and {Labarga}, F. and {Lafarga}, M. and {Llamas}, M. and {Lamp{\'o}n}, M. and {Lara}, L.~M. and {Launhardt}, R. and {L{\'a}zaro}, F.~J. and {Lodieu}, N. and {L{\'o}pez del Fresno}, M. and {L{\'o}pez-Puertas}, M. and {L{\'o}pez Salas}, J.~F. and {L{\'o}pez-Santiago}, J. and {Mag{\'a}n Madinabeitia}, H. and {Mall}, U. and {Mancini}, L. and {Mandel}, H. and {Marfil}, E. and {Mar{\'\i}n Molina}, J.~A. and {Maroto Fern{\'a}ndez}, D. and {Mart{\'\i}n}, E.~L. and {Mart{\'\i}n-Fern{\'a}ndez}, P. and {Mart{\'\i}n-Ruiz}, S. and {Marvin}, C.~J. and {Mirabet}, E. and {Monta{\~n}{\'e}s-Rodr{\'\i}guez}, P. and {Montes}, D. and {Moreno-Raya}, M.~E. and {Nagel}, E. and {Naranjo}, V. and {Narita}, N. and {Nortmann}, L. and {Nowak}, G. and {Ofir}, A. and {Oshagh}, M. and {Panduro}, J. and {Parviainen}, H. and {Pascual}, J. and {Passegger}, V.~M. and {Pavlov}, A. and {Pedraz}, S. and {P{\'e}rez-Calpena}, A. and {P{\'e}rez Medialdea}, D. and {Perger}, M. and {Perryman}, M.~A.~C. and {Rabaza}, O. and {Ram{\'o}n Ballesta}, A. and {Rebolo}, R. and {Redondo}, P. and {Reffert}, S. and {Reinhardt}, S. and {Rhode}, P. and {Rix}, H. -W. and {Rodler}, F. and {Rodr{\'\i}guez Trinidad}, A. and {Rosich}, A. and {Sadegi}, S. and {S{\'a}nchez-Blanco}, E. and {S{\'a}nchez Carrasco}, M.~A. and {S{\'a}nchez-L{\'o}pez}, A. and {Sanz-Forcada}, J. and {Sarkis}, P. and {Sarmiento}, L.~F. and {Sch{\"a}fer}, S. and {Schmitt}, J.~H.~M.~M. and {Sch{\"o}fer}, P. and {Schweitzer}, A. and {Seifert}, W. and {Shulyak}, D. and {Solano}, E. and {Sota}, A. and {Stahl}, O. and {Stock}, S. and {Strachan}, J.~B.~P. and {Stuber}, T. and {St{\"u}rmer}, J. and {Su{\'a}rez}, J.~C. and {Tabernero}, H.~M. and {Tala Pinto}, M. and {Trifonov}, T. and {Veredas}, G. and {Vico Linares}, J.~I. and {Vilardell}, F. and {Wagner}, K. and {Wolthoff}, V. and {Xu}, W. and {Yan}, F. and {Zapatero Osorio}, M.~R.},
        title = "{The CARMENES search for exoplanets around M dwarfs. Two temperate Earth-mass planet candidates around Teegarden's Star}",
      journal = {\aap},
         year = 2019,
        month = jul,
       volume = {627},
          eid = {A49},
        pages = {A49},
          doi = {10.1051/0004-6361/201935460},
archivePrefix = {arXiv},
       eprint = {1906.07196},
 primaryClass = {astro-ph.EP},
       adsurl = {https://ui.adsabs.harvard.edu/abs/2019A&A...627A..49Z}
}

@ARTICLE{Zeng+2016,
       author = {{Zeng}, Li and {Sasselov}, Dimitar D. and {Jacobsen}, Stein B.},
        title = "{Mass-Radius Relation for Rocky Planets Based on PREM}",
      journal = {\apj},
         year = 2016,
        month = mar,
       volume = {819},
       number = {2},
          eid = {127},
        pages = {127},
          doi = {10.3847/0004-637X/819/2/127},
archivePrefix = {arXiv},
       eprint = {1512.08827},
 primaryClass = {astro-ph.EP},
       adsurl = {https://ui.adsabs.harvard.edu/abs/2016ApJ...819..127Z}
}

@ARTICLE{LIFE12,
       author = {{Angerhausen}, Daniel and {Pidhorodetska}, Daria and {Leung}, Michaela and {Hansen}, Janina and {Alei}, Eleonora and {Dannert}, Felix and {Kammerer}, Jens and {Quanz}, Sascha P. and {Schwieterman}, Edward W. and {The LIFE initiative}},
        title = "{Large Interferometer For Exoplanets (LIFE). XII. The Detectability of Capstone Biosignatures in the Mid-infrared{\textemdash}Sniffing Exoplanetary Laughing Gas and Methylated Halogens}",
      journal = {\aj},
         year = 2024,
        month = mar,
       volume = {167},
       number = {3},
          eid = {128},
        pages = {128},
          doi = {10.3847/1538-3881/ad1f4b},
       adsurl = {https://ui.adsabs.harvard.edu/abs/2024AJ....167..128A}
}

@ARTICLE{LIFE8,
       author = {{Angerhausen}, Daniel and {Ottiger}, Maurice and {Dannert}, Felix and {Miguel}, Yamila and {Sousa-Silva}, Clara and {Kammerer}, Jens and {Menti}, Franziska and {Alei}, Eleonora and {Konrad}, Bj{\"o}rn S. and {Wang}, Haiyang S. and {Quanz}, Sascha P. and {LIFE Collaboration}},
        title = "{Large Interferometer for Exoplanets: VIII. Where Is the Phosphine? Observing Exoplanetary PH$_{3}$ with a Space-Based Mid-Infrared Nulling Interferometer}",
      journal = {Astrobiology},
         year = 2023,
        month = feb,
       volume = {23},
       number = {2},
        pages = {183-194},
          doi = {10.1089/ast.2022.0010},
archivePrefix = {arXiv},
       eprint = {2211.04975},
 primaryClass = {astro-ph.EP},
       adsurl = {https://ui.adsabs.harvard.edu/abs/2023AsBio..23..183A}
}

@ARTICLE{LIFE5,
       author = {{Alei}, Eleonora and {Konrad}, Bj{\"o}rn S. and {Angerhausen}, Daniel and {Grenfell}, John Lee and {Molli{\`e}re}, Paul and {Quanz}, Sascha P. and {Rugheimer}, Sarah and {Wunderlich}, Fabian and {LIFE Collaboration}},
        title = "{Large Interferometer For Exoplanets (LIFE). V. Diagnostic potential of a mid-infrared space interferometer for studying Earth analogs}",
      journal = {\aap},
         year = 2022,
        month = sep,
       volume = {665},
          eid = {A106},
        pages = {A106},
          doi = {10.1051/0004-6361/202243760},
archivePrefix = {arXiv},
       eprint = {2204.10041},
 primaryClass = {astro-ph.EP},
       adsurl = {https://ui.adsabs.harvard.edu/abs/2022A&A...665A.106A}
}

@ARTICLE{LIFE4,
       author = {{Hansen}, Jonah T. and {Ireland}, Michael J. and {LIFE Collaboration}},
        title = "{Large Interferometer For Exoplanets (LIFE). IV. Ideal kernel-nulling array architectures for a space-based mid-infrared nulling interferometer}",
      journal = {\aap},
         year = 2022,
        month = aug,
       volume = {664},
          eid = {A52},
        pages = {A52},
          doi = {10.1051/0004-6361/202243107},
archivePrefix = {arXiv},
       eprint = {2201.04891},
 primaryClass = {astro-ph.IM},
       adsurl = {https://ui.adsabs.harvard.edu/abs/2022A&A...664A..52H}
}

@ARTICLE{LIFE3,
       author = {{Konrad}, B.~S. and {Alei}, E. and {Quanz}, S.~P. and {Angerhausen}, D. and {Carri{\'o}n-Gonz{\'a}lez}, {\'O}. and {Fortney}, J.~J. and {Grenfell}, J.~L. and {Kitzmann}, D. and {Molli{\`e}re}, P. and {Rugheimer}, S. and {Wunderlich}, F. and {LIFE Collaboration}},
        title = "{Large Interferometer For Exoplanets (LIFE). III. Spectral resolution, wavelength range, and sensitivity requirements based on atmospheric retrieval analyses of an exo-Earth}",
      journal = {\aap},
         year = 2022,
        month = aug,
       volume = {664},
          eid = {A23},
        pages = {A23},
          doi = {10.1051/0004-6361/202141964},
archivePrefix = {arXiv},
       eprint = {2112.02054},
 primaryClass = {astro-ph.EP},
       adsurl = {https://ui.adsabs.harvard.edu/abs/2022A&A...664A..23K}
}

@ARTICLE{LIFE2,
       author = {{Dannert}, Felix A. and {Ottiger}, Maurice and {Quanz}, Sascha P. and {Laugier}, Romain and {Fontanet}, Emile and {Gheorghe}, Adrian and {Absil}, Olivier and {Dandumont}, Colin and {Defr{\`e}re}, Denis and {Gasc{\'o}n}, Carlos and {Glauser}, Adrian M. and {Kammerer}, Jens and {Lichtenberg}, Tim and {Linz}, Hendrik and {Loicq}, Jer{\^o}me and {LIFE Collaboration}},
        title = "{Large Interferometer For Exoplanets (LIFE). II. Signal simulation, signal extraction, and fundamental exoplanet parameters from single-epoch observations}",
      journal = {\aap},
         year = 2022,
        month = aug,
       volume = {664},
          eid = {A22},
        pages = {A22},
          doi = {10.1051/0004-6361/202141958},
archivePrefix = {arXiv},
       eprint = {2203.00471},
 primaryClass = {astro-ph.EP},
       adsurl = {https://ui.adsabs.harvard.edu/abs/2022A&A...664A..22D}
}

@ARTICLE{LIFE1,
       author = {{Quanz}, S.~P. and {Ottiger}, M. and {Fontanet}, E. and {Kammerer}, J. and {Menti}, F. and {Dannert}, F. and {Gheorghe}, A. and {Absil}, O. and {Airapetian}, V.~S. and {Alei}, E. and {Allart}, R. and {Angerhausen}, D. and {Blumenthal}, S. and {Buchhave}, L.~A. and {Cabrera}, J. and {Carri{\'o}n-Gonz{\'a}lez}, {\'O}. and {Chauvin}, G. and {Danchi}, W.~C. and {Dandumont}, C. and {Defr{\'e}re}, D. and {Dorn}, C. and {Ehrenreich}, D. and {Ertel}, S. and {Fridlund}, M. and {Garc{\'\i}a Mu{\~n}oz}, A. and {Gasc{\'o}n}, C. and {Girard}, J.~H. and {Glauser}, A. and {Grenfell}, J.~L. and {Guidi}, G. and {Hagelberg}, J. and {Helled}, R. and {Ireland}, M.~J. and {Janson}, M. and {Kopparapu}, R.~K. and {Korth}, J. and {Kozakis}, T. and {Kraus}, S. and {L{\'e}ger}, A. and {Leedj{\"a}rv}, L. and {Lichtenberg}, T. and {Lillo-Box}, J. and {Linz}, H. and {Liseau}, R. and {Loicq}, J. and {Mahendra}, V. and {Malbet}, F. and {Mathew}, J. and {Mennesson}, B. and {Meyer}, M.~R. and {Mishra}, L. and {Molaverdikhani}, K. and {Noack}, L. and {Oza}, A.~V. and {Pall{\'e}}, E. and {Parviainen}, H. and {Quirrenbach}, A. and {Rauer}, H. and {Ribas}, I. and {Rice}, M. and {Romagnolo}, A. and {Rugheimer}, S. and {Schwieterman}, E.~W. and {Serabyn}, E. and {Sharma}, S. and {Stassun}, K.~G. and {Szul{\'a}gyi}, J. and {Wang}, H.~S. and {Wunderlich}, F. and {Wyatt}, M.~C. and {LIFE Collaboration}},
        title = "{Large Interferometer For Exoplanets (LIFE). I. Improved exoplanet detection yield estimates for a large mid-infrared space-interferometer mission}",
      journal = {\aap},
         year = 2022,
        month = aug,
       volume = {664},
          eid = {A21},
        pages = {A21},
          doi = {10.1051/0004-6361/202140366},
archivePrefix = {arXiv},
       eprint = {2101.07500},
 primaryClass = {astro-ph.EP},
       adsurl = {https://ui.adsabs.harvard.edu/abs/2022A&A...664A..21Q}
}

@ARTICLE{Kammerer+2022,
       author = {{Kammerer}, Jens and {Quanz}, Sascha P. and {Dannert}, Felix and {LIFE Collaboration}},
        title = "{Large Interferometer For Exoplanets (LIFE). VI. Detecting rocky exoplanets in the habitable zones of Sun-like stars}",
      journal = {\aap},
         year = 2022,
        month = dec,
       volume = {668},
          eid = {A52},
        pages = {A52},
          doi = {10.1051/0004-6361/202243846},
archivePrefix = {arXiv},
       eprint = {2210.01782},
 primaryClass = {astro-ph.EP},
       adsurl = {https://ui.adsabs.harvard.edu/abs/2022A&A...668A..52K}
}

@BOOK{2021pdaa.book.....N,
       author = {National Academies of Sciences, Engineering and Medicine},
        title = "{Pathways to Discovery in Astronomy and Astrophysics for the 2020s}",
         year = 2021,
          doi = {10.17226/26141},
       adsurl = {https://ui.adsabs.harvard.edu/abs/2021pdaa.book.....N}
}

\begin{appendix}

\section{Surface temperatures along the equator}
\label{ap:tsurf_1D}

\begin{figure*}
\centering
\includegraphics[width=\hsize]{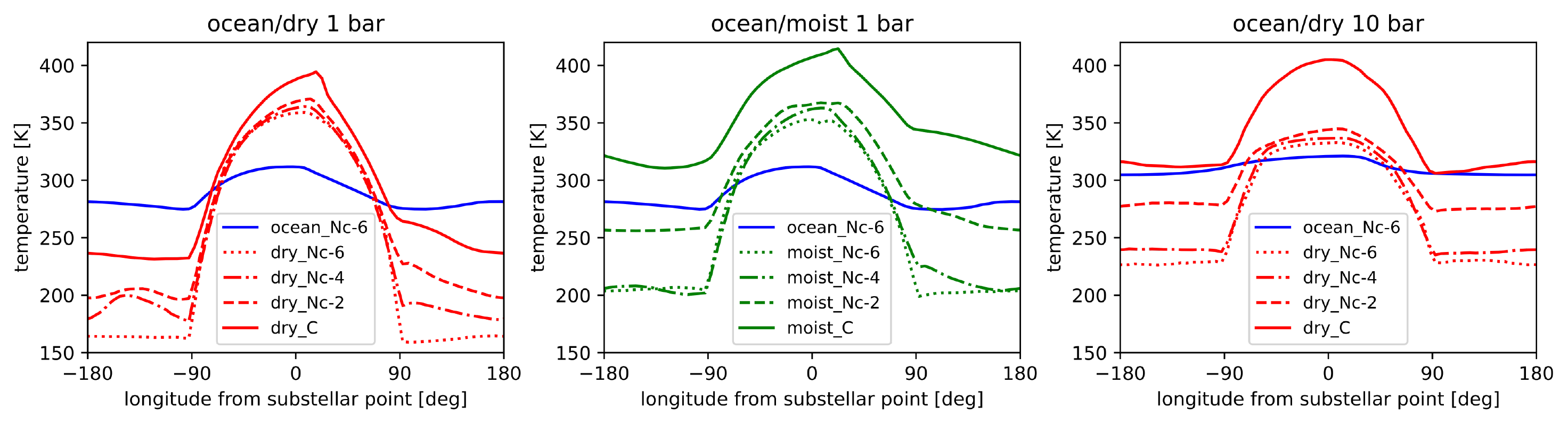}
\caption{Equatorial temperature structures for all scenarios considered in this study. Temperatures are averaged over latitudes within $\pm 10^{\circ}$.}
\label{fig:tsurf_1D}
\end{figure*}

To present the longitudinal structure to the reader more quantitatively, Figure \ref{fig:tsurf_1D} shows the equatorial temperatures as a function of longitude, averaged over latitudes within $\pm 10^{\circ}$.

\section{Thermal emission spectra based on a 2D toy model}
\label{ap:2Dtoymodel}

\begin{figure}
\centering
\includegraphics[width=0.3\hsize]{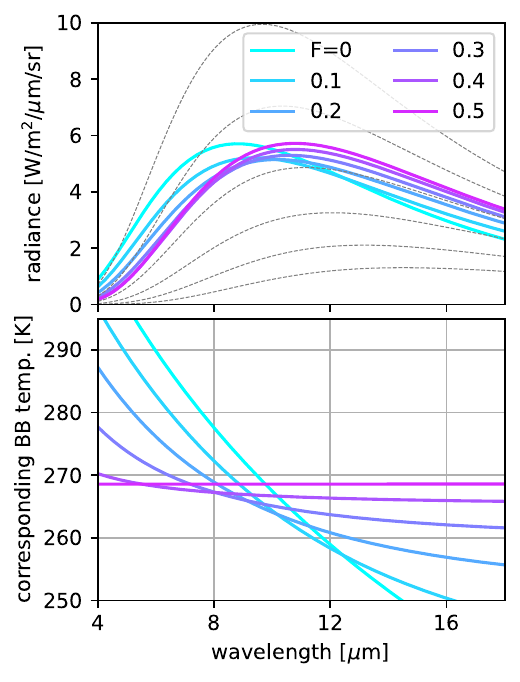}
\caption{Thermal emission spectra at quadrature of a toy planet with varying heat re-distribution efficiency. The surface temperature map is determined using the equations in \citet{KreidbergLoeb2016}. No atmospheric opacity is included. The black dotted lines correspond to blackbody emission at 200~K, 220~K, 240~K, 260~K, 280~K, and 300~K from bottom to top. }
\label{fig:snapshot_toy}
\end{figure}

The effect of the horizontal temperature gradient on the thermal emission continuum can be demonstrated with a 2D toy model. 
Here we employ the 2D model of \citet{KreidbergLoeb2016}. 
Assuming the incident flux of Teegarden's Star b and albedo 0.2, we constructed the surface temperature map with varying heat re-distribution efficiency ($F$ in \citet{KreidbergLoeb2016}), and computed the corresponding disk-averaged thermal emission spectra as viewed in a face-on configuration ($i=0^{\circ}$). The spectra are shown in Figure \ref{fig:snapshot_toy}. 
The large horizontal temperature gradient manifests itself as a decreasing brightness temperature towrad longer wavelengths. 

\end{appendix}

\end{document}